\documentclass[letterpaper,12pt,authoryear]{elsarticle}
\pdfoutput=1

\usepackage{natbib}
\usepackage{amssymb}
\usepackage{amsmath}
\usepackage{lineno}
\usepackage{url}

\bibpunct[ ]{(}{)}{,}{a}{}{,}

\makeatletter
\def\ps@pprintTitle{%
     \let\@oddhead\@empty
     \let\@evenhead\@empty
     \def\@oddfoot{\footnotesize\itshape
       Preprint accepted for publication in Icarus
       \hfill November 22, 2016}%
     \let\@evenfoot\@oddfoot}

\let\@evenhead\@empty
\def\@oddhead{\footnotesize\itshape Mid-infrared spectra of comet nuclei\hfill M.~S.~P.~Kelley et al.}
\let\@evenfoot\@empty
\def\@oddfoot{\footnotesize\itshape Preprint accepted for publication in Icarus\hfill {\normalfont \thepage} \hfill November 22, 2016}
\makeatother
\usepackage{tabularx}

\usepackage{textcomp}

\usepackage{hyperref}

\newcommand\micron{\mbox{\textmu m}}%
\newcommand\arcsec{\mbox{$^{\prime\prime}$}}%
\newcommand\arcdeg{\mbox{$^\circ$}}%
\newcommand\arcmin{\mbox{$^\prime$}}%
\newcommand\degr{\arcdeg}%
\providecommand\mjysr{MJy\,sr$^{-1}$}

\newcommand\wm{W\,m$^{-2}$}

\newcommand\mks{J\,K$^{-1}$\,m$^{-2}$\,s$^{-1/2}$} % thermal inertia
\newcommand\coo{CO$_2$}

\newcommand\sst{\textit{Spitzer Space Telescope}}
\newcommand\spitzer{\textit{Spitzer}}

\newcommand\rosetta{\textit{Rosetta}}
\newcommand\di{\textit{Deep Impact}}

\newcommand\iso{\textit{ISO}}
\newcommand\iraslong{\textit{Infrared Astronomical Satellite}}

\newcommand\nodata{-}%

\journal{Icarus}

\begin{document}
\begin{frontmatter}

\title{Mid-infrared spectra of comet nuclei}

\author[umd]{Michael S. P. Kelley\corref{cor}}
\cortext[cor]{Corresponding author}
\ead{msk@astro.umd.edu}

\author[umn]{Charles E. Woodward}

\author[umn]{Robert D. Gehrz}

\author[sofia]{William T. Reach}

\author[ucsd]{David E. Harker}

\address[umd]{Department of Astronomy, University of Maryland, College
  Park, MD 20742-2421, USA}

\address[umn]{Minnesota Institute for Astrophysics, School of Physics
  and Astronomy, 116 Church Street S. E., University of Minnesota,
  Minneapolis, MN 55455, USA}

\address[sofia]{Universities Space Research Corporation, Stratospheric
  Observatory for Infrared Astronomy, MS 232-11, NASA Ames Research
  Center, Moffett Field, CA 94035, USA}

\address[ucsd]{Center for Astrophysics and Space Sciences, University
  of California, San Diego, 9500 Gilman Drive, La Jolla, CA
  92093-0424, USA}

\begin{abstract}
  Comet nuclei and D-type asteroids have several similarities at optical and near-IR wavelengths, including near-featureless red reflectance spectra, and low albedos.  Mineral identifications based on these characteristics are fraught with degeneracies, although some general trends can be identified.  In contrast, spectral emissivity features in the mid-infrared provide important compositional information that might not otherwise be achievable.  Jovian Trojan D-type asteroids have emissivity features strikingly similar to comet comae, suggesting that they have the same compositions and that the surfaces of the Trojans are highly porous.  However, a direct comparison between a comet and asteroid surface has not been possible due to the paucity of spectra of comet nuclei at mid-infrared wavelengths.  We present 5--35~\micron{} thermal emission spectra of comets 10P/Tempel~2, and 49P/Arend-Rigaux observed with the Infrared Spectrograph on the \sst{}.  Our analysis reveals no evidence for a coma or tail at the time of observation, suggesting the spectra are dominated by the comet nucleus.  We fit each spectrum with the near-Earth asteroid thermal model (NEATM) and find sizes in agreement with previous values.  However, the NEATM beaming parameters of the nuclei, 0.74 to 0.83, are systematically lower than the Jupiter-family comet population mean of 1.03$\pm$0.11, derived from 16- and 22-\micron{} photometry.  We suggest this may be either an artifact of the spectral reduction, or the consequence of an emissivity low near 16~\micron. When the spectra are normalized by the NEATM model, a weak 10-\micron{} silicate plateau is evident, with a shape similar to those seen in mid-infrared spectra of D-type asteroids.  A silicate plateau is also evident in previously published \spitzer{} spectra of the nucleus of comet 9P/Tempel~1.  We compare, in detail, these comet nucleus emission features to those seen in spectra of the Jovian Trojan D-types (624) Hektor, (911) Agamemnon, and (1172) Aneas, as well as those seen in the spectra of seven comet comae.  The comet comae present silicate features with two distinct shapes, either trapezoidal, or more rounded, the latter apparently due to enhanced emission near 8 to 8.5~\micron.  The surfaces of Tempel~2, Arend-Rigaux, and Hektor best agree with the comae that present trapezoidal features, furthering the hypothesis that the surfaces of these targets must have high porosities in order to exhibit a spectrum similar to a comet coma.  An emissivity minimum at 15~\micron, present in the spectra of Tempel~2, Arend-Rigaux, Hektor, and Agamemnon, is also described, the origin of which remains unidentified.  The compositional similarity between D-type asteroids and comets is discussed, and our data supports the hypothesis that they have similar origins in the early Solar System.

% revision
\end{abstract}

\end{frontmatter}

%\linenumbers
\section{Introduction}\label{sec:intro}

At the heart of every comet is a nucleus composed of dust, ice, and gas.  Under certain circumstances the nucleus can be observed with remote sensing techniques.  Reflectance spectroscopy of comet nuclei has revealed spectral similarities akin to certain asteroid taxonomic classes \citep{tholen84, bus02-tax, demeo09}.  Classification in these taxonomy schemes is typically based on principal component analysis of color photometry or spectra, including tests for the presence or absence of certain spectral features and consideration of overall spectral slope.  There are approximately two dozen classes and sub-classes of asteroids.  The spectra of some cometary nuclei resemble those asteroid classes with red-sloped reflectance spectra lacking strong spectral features.

Comet nuclei tend to have linear reflectance spectra, with red slopes up to $\approx$20\% in flux per 0.1~\micron{} in the optical (see \citealt{jewitt02}, \citealt{campins07}, and references therein).  D-type asteroids have similar slopes \citep[7--14\% per 0.1~\micron{}][]{fitzsimmons94}.  Six comets studied by \citet{demeo08} were classified as D-, T-, and X-type taxonomies (in a general sense, T-type spectra are intermediate to D-type and the more shallowly sloped X-type spectra).  By contrast, spectroscopy of comets 133P/Elst-Pizarro and 176P/LINEAR \citep{licandro11} reveals B-type asteroid class spectra, i.e., blue-to-neutrally sloped in the optical, with an apparent near-UV absorption feature.  Both of these latter comets have orbits and spectral classifications similar to the Themis collisional family of the outer main asteroid belt \citep{hsieh06, licandro11}, suggesting they have similar origins.

The paucity of features in the reflectance spectra of comets, and D-, T-, and X-type asteroids limits surface compositional interpretations.  \citet{campins06} compared the nucleus spectrum of comet 162P/Siding Spring to those of chondritic meteorites, but found no appropriate matches.  Instead, synthetic mixtures of amorphous carbon, pyroxenes, and/or organic compounds produced spectra similar to that of the comet surface.  However, unambiguous and unique conclusions regarding the nucleus surface composition of 162P was not possible due to degeneracies in the models used to generate the synthetic spectra.  \citet{emery11} also used synthetic mixtures of various silicates and carbon to reproduce the observed near-IR (0.7--2.5~\micron) spectra of 58 Jovian Trojan asteroids with D- and X-type classifications.  They concluded that the Trojan population has a bimodal composition, but refrained from identifying specific minerals.

\citet{hiroi01} examined samples of the Tagish Lake meteorite, an aqueously altered carbonaceous chondrite, and demonstrated that these meteorite fragments have optical to near-IR spectra similar to those of T- and D-type asteroids.  This meteorite also has a 3-\micron{} absorption band indicative of hydration, a feature that is present in spectra of the T-type asteroid (308) Polyxo \citep{hiroi03, takir12}, but absent from the D-type population \citep{jones90, hiroi01, emery03}.  Clearly, without the 3-\micron{} study, an incorrect assumption would have resulted, i.e., that all three samples (Tagish Lake, D-types, and T-types) have similar compositions.  These examples demonstrate the difficulty in understanding the composition of spectrally bland surfaces over limited wavelength regimes.

Mid-infrared spectroscopy ($\lambda>5$~\micron) of asteroids, dominated by thermal emission from the surface, is also sensitive to surface composition.  Thermal emission spectra can present diagnostic spectral features from asteroids that would otherwise be featureless in reflectance \citep{dotto04, marchis12}.  Considering, again, the Tagish Lake meteorite, \citet{vernazza13} find an excellent match in the mid-infrared with asteroid (368) Haidea but poor agreement with (617) Patroclus, despite the fact that both asteroids have nearly identical near-infrared reflectance spectra.

\citet{emery06} observed three Jovian Trojans (D-types) with the \sst{}, revealing an emissivity plateau near 10~\micron{} and a broad rise in the 18--28~\micron{} spectral region, including narrow and subtle spectral features at 20 and 34~\micron.  The gross shapes and locations of the features were remarkably similar to those found in the spectra of the comae around comets C/1995 O1 (Hale-Bopp) and 29P/Schwass\-mann-Wach\-mann~1 \citep{crovisier97-science, stansberry04}.  \citet{emery06} attribute the spectral similarity to the presence of small silicate grains on the surfaces of these asteroids.  They posit that the grains are suspended in a transparent matrix or porous structure \citep[i.e., the ``fairy-castle'' structures described by][]{hapke63}, enabling the silicates to emit without interference from nearby absorbing or scattering grains, much as an isolated grain would in a comet coma.

Comets 10P/Tempel~2 and 49P/Arend-Rigaux are Jupiter-family comets (JFCs) well known for having large nuclei and weak or negligible coma activity during their approach to perihelion.  Thus, the bare nuclei of these comets can be studied and contrasted with asteroid spectra at optical and mid-infrared wavelengths.  Here, we present \sst{} spectroscopy and imaging of the nuclei of comets Tempel~2 and Arend-Rigaux.  We also present a spectrum of the nucleus of comet 9P/Tempel~1 from \citet{lisse05-nucleus}.  In Section~\ref{sec:obs}, we summarize the known properties of each comet, the observations, and our reduction methods.  In Section~\ref{sec:model}, we describe the near-Earth asteroid thermal model (NEATM) of \citet{harris98} as it applies to comets.  In Section~\ref{sec:results}, we apply the NEATM to our data, and present derived thermal parameters and emissivity spectra.  We compare the spectra of these three comet nuclei to spectra of Jovian D-type asteroids, and active comets.  Our conclusions are summarized in Section~\ref{sec:summary}.

\section{Targets, Observations, and Reduction}\label{sec:obs}
Comet Tempel~2, discovered on 4 July 1873, is a JFC with a period of 5.4~yr and a perihelion distance of 1.42~au \citep[comet discovery and orbital data are from JPL HORIZONS;][]{giorgini96}.  The comet's coma is faint, and appreciable activity is usually not seen until 100 days before perihelion at heliocentric distances, $r_h$, $\lesssim$1.75~au \citep{sekanina79}.  \citet{jewitt89} found that the coma onset began near $2.1$~au in 1988 (160 days pre-perihelion).  However, observations of the comet by \citet{reach13} at 3.2~au in 2009 (340 days pre-perihelion) revealed a faint gas coma (\coo{} or CO) without a clear dust counterpart.  \citet{ahearn89} presented simultaneous optical and infrared photometry of Tempel~2.  They found a moderate coma contribution ($\sim25$\%) to the optical light, but that the coma was undetectable at the 10\% level in the IR.  The optical and IR light curves are in phase, leading \citet{ahearn89} to conclude that most of the radiation comes from a rotating nucleus with a maximum effective radius of 5.9~km \citep[using the standard thermal model for asteroids;][]{lebofsky86}, an axial ratio $>1.9$, and a geometric albedo, $A_p$, of 0.022 in the optical with a reflectivity gradient of 15\% per 0.1~\micron.  \citet{jewitt89} later revised the albedo to $0.024\pm0.005$.  Comet Tempel~2 has an extensive dust trail, stretching out to 60\degr{} of mean anomaly from the nucleus as observed by the \iraslong{} \citep{sykes90}.

Comet Arend-Rigaux, discovered on 5 February 1951, has a period of 6.7~yr and a perihelion distance of 1.43~au.  Similar to the behavior of Tempel~2, Arend-Rigaux also has a delayed onset of pre-perihelion activity, and weak overall activity.  Activity is not readily apparent in the comet's total magnitude until 100 days before perihelion\footnote{For example, see Seichii Yoshida's compiled lightcurves: \url{http://www.aerith.net/comet/catalog/0049P/index.html}}. \citet{tokunaga85} observed the comet in the infrared in January 1985 ($r_h=$1.54~au, $\Delta=$0.57~au), and found a nucleus-dominated aperture profile with no evidence for a dust coma.  They estimate an effective radius of 4.4--5.1~km by fitting the thermal emission with a 310~K blackbody sphere. \citet{brooke86} interpreted their own infrared observations taken about a month later with the thermal model of \citet{morrison79} and derived an effective radius of $5.1\pm1.1$~km with $A_p=0.02\pm0.01$.  The low albedo and size was confirmed by \citet{veeder87} and \citet{millis88}; the latter team estimated the nucleus to have an equivalent prolate spheroidal shape of $13\times8\times8$~km ($5.2\pm0.2$~km effective radius at their observed maximum) and $A_p=0.028\pm0.005$.  Several investigators pointed out the similarity of the reflectance spectra of Arend-Rigaux's nucleus to those of D-type asteroids \citep{brooke86, veeder87, millis88}.  \citet{luu93} describes optical spectra of this comet at 2.2--2.4~au with a reflectivity gradient of 9--11\% per 0.1~\micron.

Comet Tempel~1, discovered 3 April 1867, has a period of 5.6~yr and a perihelion distance of 1.54 au.  In contrast to Tempel~2 and Arend-Rigaux, Tempel~1 has a typical activity level for its size.  \citet{lisse05-nucleus} estimated the active fraction, the ratio of the surface area required to match the water production rate to the surface area of the nucleus, to be 9$\pm$2\%.  This fraction is much larger than the estimates for Tempel~2 and Arend-Rigaux approximately 1\% and $<$1\%, respectively \citep{millis88, ahearn89}. Based on the compilation of photometry by \citet{meech05-ssr}, the first signs of activity of comet Tempel~1 occur between 600 and 400 days before perihelion (4--3~au).  \spitzer{} images of the comet at 478 and 464 days pre-perihelion (3.7~au) show no evidence for near aphelion activity \citep{lisse05-nucleus, reach07}, and neither do $R$-band images at 4.2~au post-perihelion \citep{hergenrother07}.  The comet was the target of the 2005 \textit{Deep Impact} mission to excavate material from the near surface of a nucleus \citep{ahearn05}, and the 2011 \textit{Stardust-NExT} mission to image the impact crater and other surface feature variations \citep{veverka13}.  Based on the spacecraft images and photometry, the nucleus has a mean radius of 2.8$\pm$0.1~km \citep{thomas13-tempel1}, $A_p=0.056$, and a reflectivity gradient of 12.5$\pm$1\% per 0.1~\micron{} in the optical \citep{li07}.

We and others obtained mid-infrared spectra of comets Tempel~2, Arend-Rigaux, and Tempel~1 using the Infrared Spectrograph \citep[IRS;][]{houck04} on the \sst{} \citep{werner04, gehrz07}.  A summary of the observing geometry for each comet is presented in Table~\ref{tab:obs}.  The observations were made with the short-wavelength low-resolution (SL) and the long-wavelength low-resolution (LL) modules.  Both modules have a spectral resolving power of $R=\Delta\lambda/\lambda \approx 64-128$.  The low-resolution IRS modules each use two slits to record data via two different spectral orders.  The first-order SL slit (SL1) covers 7.5--14.3~\micron{}, and the second-order slit simultaneously covers 5.1--7.6 (SL2) and 7.3--8.7~\micron{} (SL3).  Similarly, LL covers 19.9--39.9 (LL1), 13.9--21.3 (LL2), and 19.2--21.6~\micron{} (LL3).  The Tempel~1 and Tempel~2 observations used the SL1 and all LL orders, whereas the Arend-Rigaux observations used all SL and LL orders.  All three comets were observed with the 16-\micron{} peak-up array for target acquisition.  The Tempel~2 and Arend-Rigaux data were processed with the \spitzer{} Science Center's S18.18 IRS pipeline \citep{irs}.  Tempel~1 was processed with pipeline version S10.5 \citep{lisse05-nucleus}.

%%%%%%%%%%%%%%%%%%%%%%%%%%%%%%%%%%%%%%%%%%%%%%%%%%%%%%%%%%%%%%%%%%%%%%%%%%%%%%%%
% Observation details
%%%%%%%%%%%%%%%%%%%%%%%%%%%%%%%%%%%%%%%%%%%%%%%%%%%%%%%%%%%%%%%%%%%%%%%%%%%%%%%%
{\renewcommand\baselinestretch{1}
  \begin{table}
    \footnotesize
    \caption{Observation details.\label{tab:obs}}
    \begin{center}
      \makebox[\textwidth]{
      \begin{tabularx}{1.2\textwidth}{@{\extracolsep{0pt}}lllcccccc}
        \hline
        Target
        & Inst.
        & AOR Key
        & Start Time
        & $t_{exp}$
        & $T-T_p$
        & $r_h$
        & $\Delta$
        & Phase \\

        &
        &
        & (UT)
        & (s)
        & (days)
        & (AU)
        & (AU)
        & (\degr) \\
\hline

9P/Tempel 1\textsuperscript{a} & IRS             &  9347328 & 2004-03-26 03:05 & 2880                    & -466.2 & 3.70 & 3.52 & 15.6 \\

10P/Tempel 2           & IRS                     &  6616576 & 2004-06-25 19:34 &  612                    & -234.9 & 2.57 & 2.31 & 23.3 \\
10P/Tempel 2           & MIPS\textsuperscript{b} &  6614272 & 2004-07-08 10:22 &  140\textsuperscript{c} & -221.6 & 2.49 & 2.39 & 24.0 \\

49P/Arend-Rigaux       & MIPS\textsuperscript{b} & 10300416 & 2004-12-05 14:18 &   42\textsuperscript{c} &  -81.0 & 1.64 & 0.99 & 35.3 \\
49P/Arend-Rigaux       & IRS                     & 16206592 & 2006-03-07 02:31 &  900                    &  375.4 & 3.56 & 3.43 & 16.3 \\

\hline

      \end{tabularx}
}
    \end{center}

    \begin{footnotesize}
      Table columns|Inst.: \spitzer{} instrument; AOR Key: \spitzer{}
      astronomical observation request key; $t_{exp}$: total time on
      source; $T-T_p$: Difference between observation date, $T$, and
      perihelion date, $T_p$; $r_h$: heliocentric distance; $\Delta$:
      \spitzer-comet distance; Phase: Sun-comet-\spitzer{} angle.

      \textsuperscript{a} \citet{lisse05-nucleus}.  The other AORs used
      in our Tempel~1 analysis (9347584, 9348608, and 9348864) are
      taken within 21~hr of the listed observation.  The listed
      exposure time is for all four observations.

      \textsuperscript{b} \citet{reach07}.

      \textsuperscript{c} Time per pixel at the comet location.

    \end{footnotesize}
\end{table}
}

\subsection{Peak-up acquisition images}
\begin{figure}
  \includegraphics[width=\textwidth]{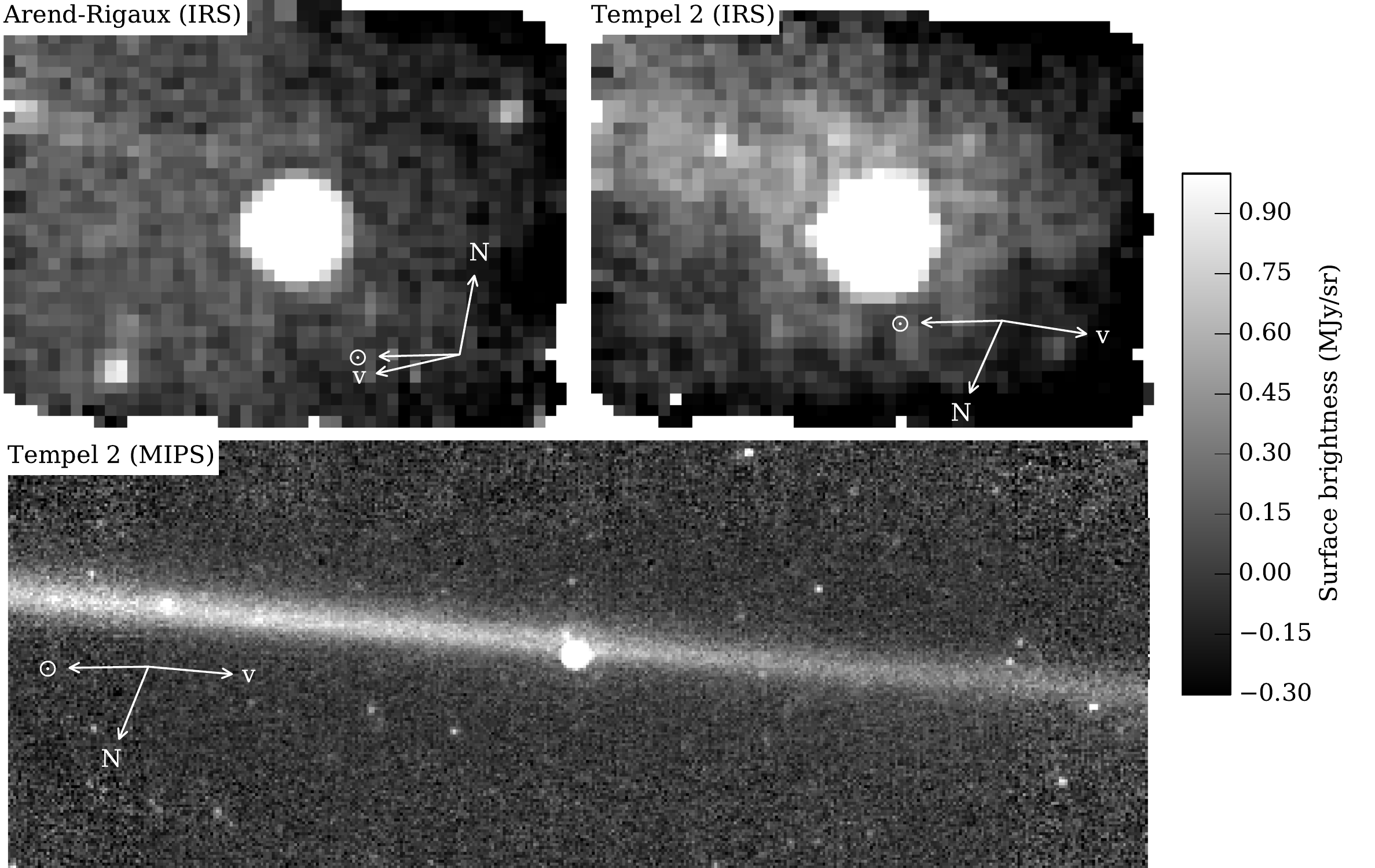}
  \caption{\spitzer/IRS 16-\micron{} peak-up acquisition images of comets 49P/Arend-Rigaux (2006 Mar 07 UT, 228,000~km $\times$ 173,000~km, top left) and 10P/Tempel~2 (2004 Jun 25 UT, 171,000~km $\times$ 130,000~km, top right), and a 24-\micron{} image of comet Tempel~2 (2004 Jul 08 UT, 1,361,000~km $\times$ 510,000~km, bottom) from \citet{reach07}.  The projected directions of celestial north (N), the Sun ($\odot$), and the comet velocity (v) are indicated by arrows.  A constant background estimate has been removed from each image and the contrast has been set to enhance the residual background.  A dust trail is revealed in both Tempel~2 images.  No dust tail or coma is seen in any image.\label{fig:images}}
\end{figure}

Images of each comet (Fig.~\ref{fig:images}) were taken with the IRS peak-up arrays in peak-up acquisition mode.  The images were used by \spitzer{} to acquire the nucleus to provide positioning of the object in the IRS slits.  The peak-up acquisition photometric calibration uncertainty is 3\% for the 16-\micron{} array \citep{irs}.  We averaged all acquisition images of Tempel~2 and Arend-Rigaux in the rest frame of each comet using the MOPEX software package \citep{makovoz05-mopex}.

\subsection{Other images}
\begin{figure}
  \includegraphics[width=\textwidth]{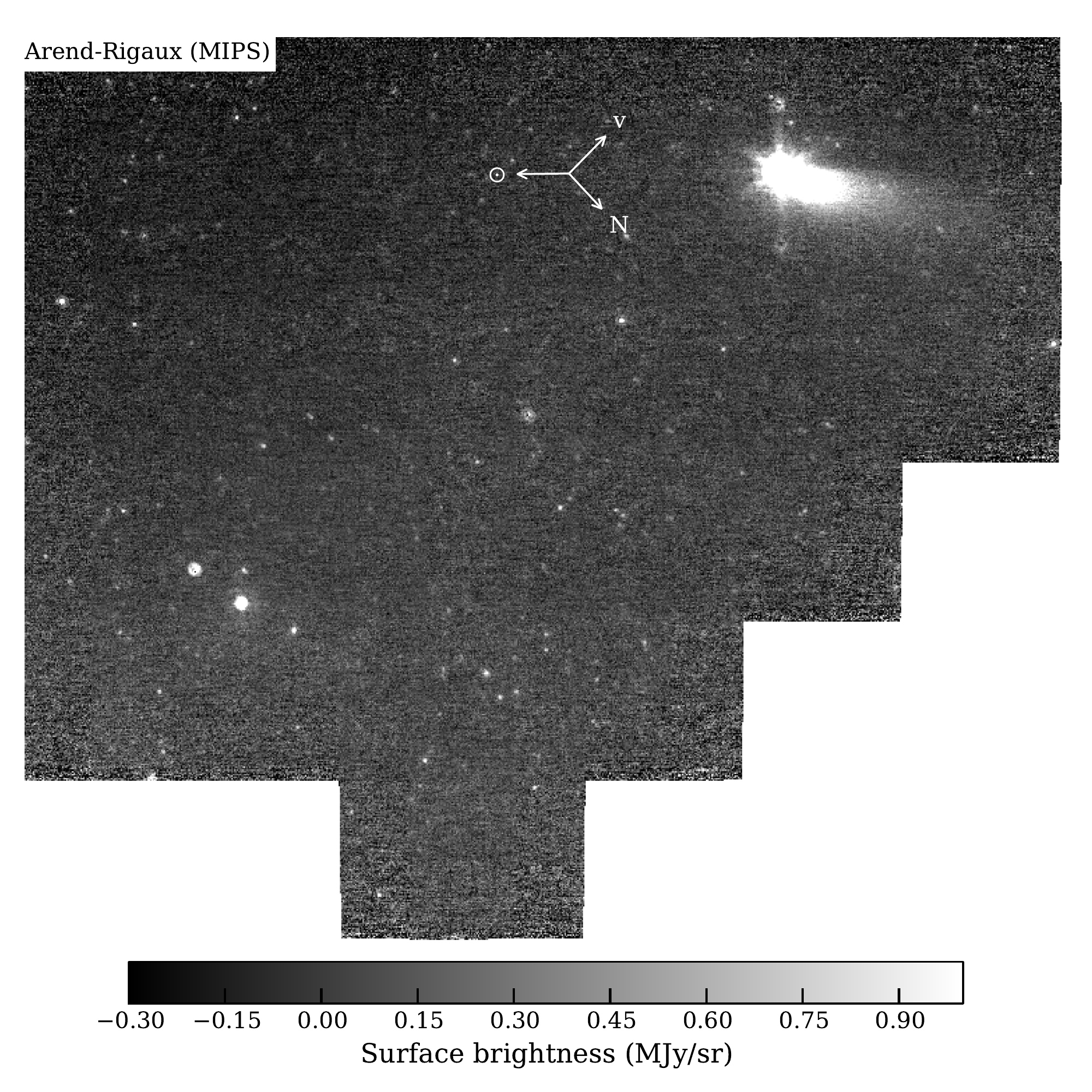}
  \caption{\spitzer/MIPS 24-\micron{} image of comet Arend-Riguax (2004 Dec 05 UT, 537,000~km $\times$ 465,000~km, bottom), taken 15 months before our \spitzer{} data, from \citet{reach07}.  A bright point source and dust tail are evident; no dust trail is seen. \label{fig:ar-image}}
\end{figure}

For additional context, we obtained images of comets Tempel~2 and Arend-Rigaux with \spitzer's Multiband Imaging Photometer (MIPS) 24-\micron{} camera \citep{rieke04} from \citet{reach07}.  This instrument uses a broad-band 24-\micron{} filter and a 128$\times$128 pixel array with a pixel scale of 2.55\arcsec{}.  The observing details for each comet at the epoch of the MIPS observations are listed in Table~\ref{tab:obs}.  The MIPS image of comet Tempel~2 was taken just two weeks after our IRS data (Fig.~\ref{fig:images}), and the MIPS image of comet Arend-Rigaux was taken 15 months before our IRS data with the comet much more active at 1.6~au from the Sun (Fig.~\ref{fig:ar-image}).

\subsection{Spectra}\label{sec:spec}
The Tempel~2 and Arend-Rigaux spectra were reduced with the \spitzer{} Science Center's SPICE reduction package\footnote{Available at the NASA/IPAC Infrared Science Archive, \url{http://irsa.ipac.caltech.edu/data/SPITZER/docs/dataanalysistools/}}.  We used the default point source aperture sizes to extract the spectra from each observation.  The point source aperture sizes are tuned to follow the wavelength dependence of the point-spread function.  Our targets appear to be point sources, and an aperture correction typically needed for observations of extended sources is not necessary.  The targets were placed at two or three positions within each slit, with three array readouts at each position.  For each wavelength, we averaged the redundant spectra together, rejecting outliers at the 2.5-$\sigma$ level or greater.

\begin{figure}
\includegraphics[width=\textwidth]{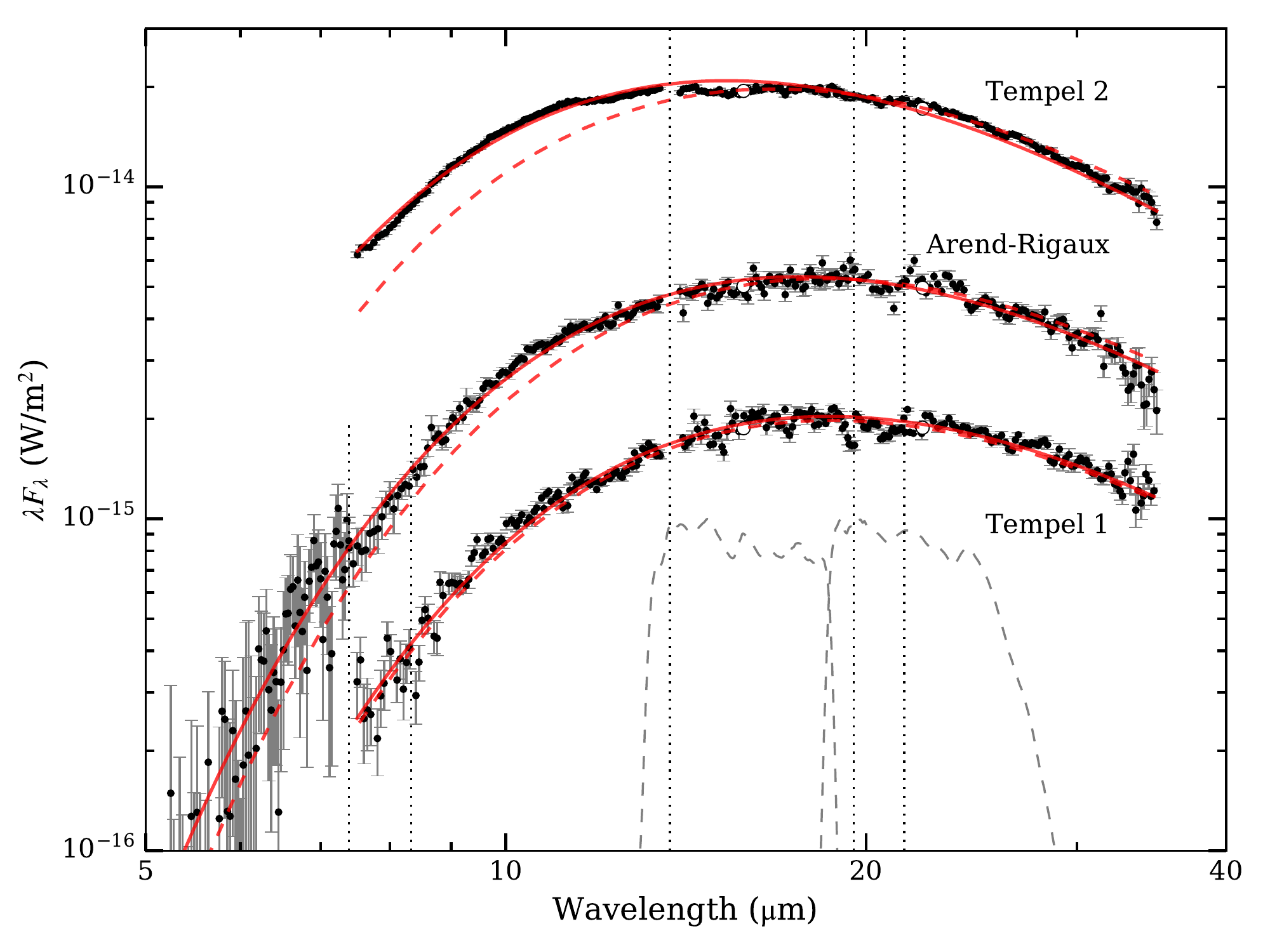}
\caption{\spitzer/IRS spectra of comets 10P/Tempel~2, 49P/Arend-Rigaux, and 9P/Tempel~1.  Tempel~1 spectra are based on the work of \citet{lisse05-nucleus}.  Vertical dotted lines mark the boundaries between different orders of the spectrometer.  The solid-lines are our best-fit nucleus thermal models.  Also shown is synthetic IRS peak-up imaging photometry (large open circles).  The IRS 16- and 22-\micron{} filter bandpasses  are given at the bottom of the plot.  The best-fit NEATM models derived from the synthetic photometry are plotted as dashed lines.}
\label{fig:spec} 
\end{figure}

The spectra of comet Tempel~2 from each slit were combined, removing order overlaps, to produce a near-continuous spectrum.  We corrected an apparent module-to-module offset, potentially caused by a misalignment of the source and slit \citep{irs}, by scaling the SL spectra by 1.12.  The scaling constant was derived by linear interpolation of the SL and LL spectra over the gap at 14~\micron{}.  The SL1 module has a data artifact of additional emission at 13.5--14.2~\micron{} known as the ``teardrop,'' most likely due to an internal reflection \citep{irs}.  Visual inspection of the raw images reveals its presence in our data.  We remove this portion of the spectrum, and present the final spectrum of comet Tempel~2 in Fig.~\ref{fig:spec} and in the on-line supplementary material.

The spectra of Arend-Rigaux were combined in a similar manner as was done for the Tempel~2 data.  The SL1 order (8.4--13.4~\micron) was scaled by a factor of 1.20 to account for an apparent offset with the LL module.  The SL2 and SL3 orders (5.2--8.3~\micron) were scaled by a factor of 1.03 to match the newly scaled SL1 order.  The final spectrum of comet Arend-Rigaux is presented in Fig.~\ref{fig:spec} and in the on-line supplementary material.

The spectra of the nucleus of comet Tempel~1 from \citet{lisse05-nucleus} have lower signal-to-noise ratios than the Arend-Rigaux and Tempel~2 spectra.  To compensate, we combined the four brightest epochs to produce the spectrum presented in Fig.~\ref{fig:spec}.  Spectral uncertainties were not provided by \citet{lisse05-nucleus}.  We generated spectral uncertainties from the point-to-point scatter in the spectra after removing the gross spectral shape defined by Guassian smoothing.

\section{Thermal Models}\label{sec:model}
The near-Earth asteroid thermal model \citep[NEATM;][]{harris98} has successfully reproduced comet nucleus effective radii (as determined through spacecraft encounters) based on thermal emission photometry from remote sensing.  The NEATM assumes the surface of a spherical body is in instantaneous thermal equilibrium with insolation, placing the hottest spot at the sub-solar point, i.e., the surface has a thermal inertia of zero.  In general, the NEATM is conceptionally appropriate as a modeling tool if the surface has a low thermal inertia ($\lesssim$ a few hundred \mks).  The thermal properties of three spatially resolved comet surfaces have been studied and modeled.  Temperature maps of the nucleus of comets Tempel~1 and 103P/Hartley~2 derived from the \textit{Deep Impact} flyby spacecraft show that these nuclei have low to moderate thermal inertias \citep{groussin07, davidsson13-thermali, groussin13}.  \citet{davidsson13-thermali} demonstrated that the flux from Tempel~1 terrains with the highest thermal inertia values (150--200~\mks{}), deviate from 0 to 30\% from an instantaneous model.  Sub-millimeter observations of comet 67P/Churyumov-Gerasimenko with the \rosetta/MIRO instrument suggest the surface has a thermal inertia near 10--50~\mks{} \citep{gulkis15}.  Thus, comet nuclei likely have low thermal inertia surfaces.

In terms of thermal inertia, the NEATM assumptions seem appropriate for comet nuclei.  However, the shapes of our targets deviate significantly from the spherical assumption.  \citet{brown85} compare the thermal emission from ellipsoidal model asteroids to that from a spherical model.  He shows that the ellipsoids have relatively stronger 20-\micron{} emission than would otherwise be predicted, due to the presence of more facets with cooler thermal equilibrium temperatures.  In the context of the NEATM, this would be accounted for by an increase in the beaming parameter (described below).  Accounting for nucleus shape is beyond the scope of this paper.  We primarily use the NEATM to establish an effective continuum spectrum for each target, which we will use to estimate an approximate emissivity spectrum.  We report beaming parameters and effective radii in order to promote efficient comparison with other comet nucleus and asteroid observations \citep[e.g., the infrared photometric survey of][]{fernandez13}.

Within the NEATM formalism, the temperature, $T$, of a sun-lit surface element of a sphere is
\begin{equation}
  T = \left[\frac{(1 - A)S \cos{Z}}{\eta\epsilon\sigma}\right]^{1/4}
  (\mathrm{K}),
  \label{eq:neatmTemp}
\end{equation}
where $A$ is the bolometric Bond albedo (related to the geometric albedo, $A_p$, through the relation $A=A_p\,q$, where $q$ is the phase integral; \citealt{hanner81-albedo}), $S$ is the incident solar flux (e.g., \wm), $Z$ is the Sun-zenith angle, $\epsilon$ is the infrared emissivity, $\sigma$ is the Stefan-Boltzmann constant, and $\eta$ is a unitless scale factor referred to by the model as the ``beaming parameter.''  In practice, $\eta$ is a proxy for the combined effects of thermal inertia, bulk shape, and surface roughness \citep[e.g.,][]{lagerros98, delbo02, wolters08}.  Surface roughness and shape cause ``beaming'' that can lower or raise the derived $\eta$ depending on the phase angle of the observation.  Meanwhile, thermal inertia and rotation causes energy to propagate into the body and persist beyond the day/night terminator, raising the observed $\eta$ \citep{spencer89, spencer90}.  Based on a two-color (16 and 22~\micron) photometric survey of 57 comet nuclei with \spitzer{}, \citet{fernandez13} derive a JFC population averaged $\eta=1.03\pm0.11$ assuming $\epsilon=0.95$ and a geometric albedo $A_p=0.04$.  We also use $\epsilon=0.95$ and allow $\eta$ to freely vary when fitting spectra, but adopt the measured albedos for each nucleus: 0.022 for Tempel~2 \citep{ahearn89}, 0.028 for Arend-Rigaux \citep{millis88}.  The uncertainties in these albedos ($\pm$0.005) are ignored in our fits.  In order to convert from geometric albedo to Bond albedo, we set the phase parameter $G$ \citep[of the $H$, $G$ magnitude system;][]{bowell89} to the nominal asteroidal value of 0.15.  Thus, the geometric to Bond albedo conversion becomes $A = A_p (0.290 + 0.684 G)$ \citep[see][]{delbo02}.  The flux from the surface is integrated over the area visible to the observer to produce the model spectrum.

\section{Results and Analysis}\label{sec:results}
\subsection{Imaging}\label{sec:imaging}
We first examine the images of our targets to determine if they show any evidence of a dust coma, tail, or trail.  All of the IRS and MIPS images of comets Tempel~2 and Arend-Rigaux show a strong point source, and the Tempel~2 images additionally show this comet's dust trail.  The MIPS image of comet Arend-Rigaux, taken 15 months prior to our spectra when the comet was closer to the Sun, has a prominent dust tail.  This image was taken at a heliocentric distance of 1.64~au.  No dust coma or tail is seen in any of our other images.  Comet Arend-Rigaux's activity seems to have been extinguished by the time it was observed with IRS at 3.56~au.

\begin{figure}
\centering
\includegraphics[width=\textwidth]{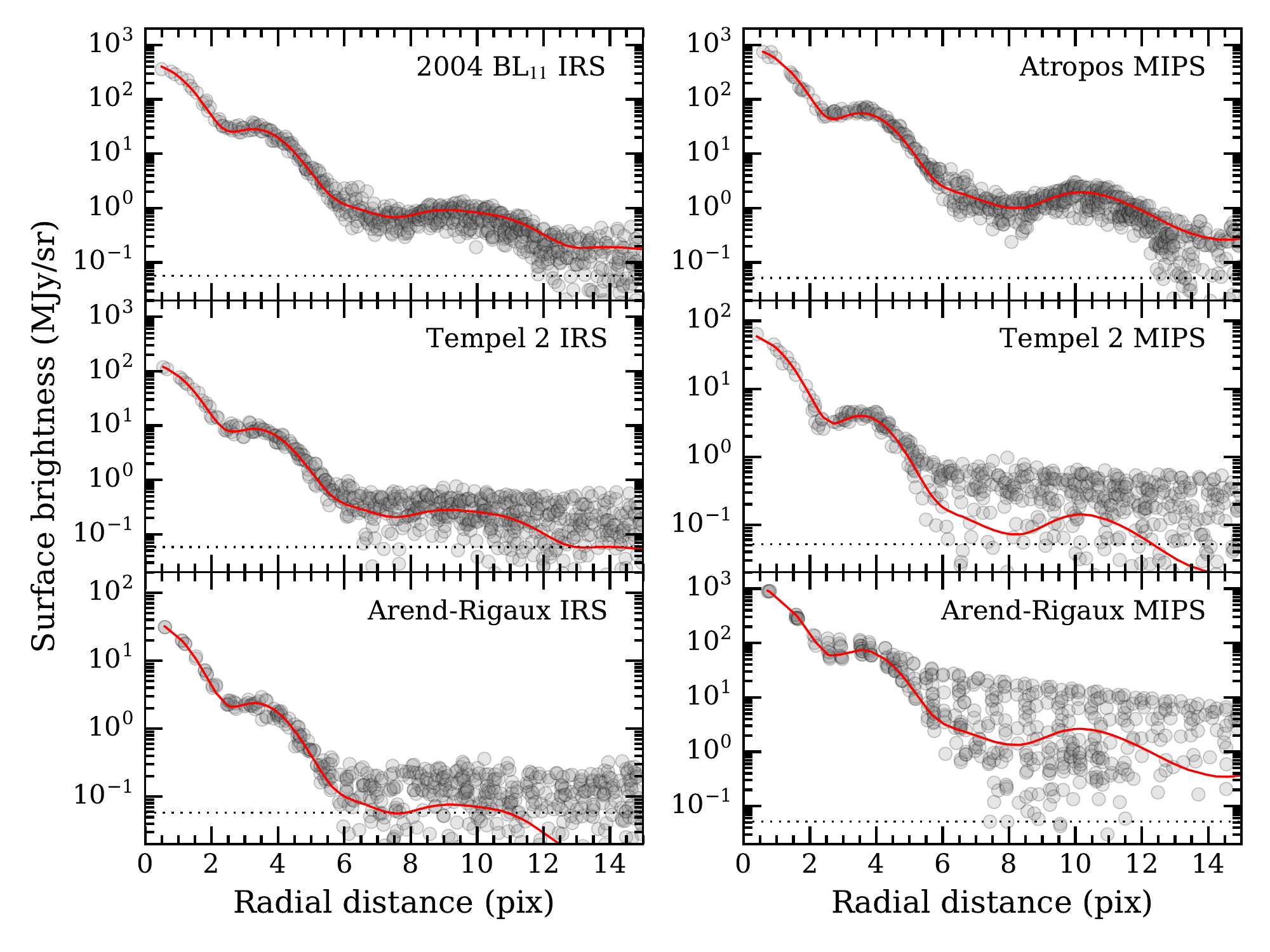}
\caption{Spatial profiles of comets 10P/Tempel~2 and 49P/Arend-Rigaux, and our example asteroids 2004~BL$_{11}$ \citep{campins09-lowq} and Atropos (Section~\ref{sec:imaging}).  A best-fit point-response function is shown for each target (solid line), as well as the 2-$\sigma$ background level (horizontal dotted line).  Extended emission, the dust trail of Tempel~2 and the coma of Arend-Rigaux, is apparent in the \spitzer/MIPS observations, as is a significant background gradient in the Arend-Rigaux \spitzer/IRS profile.  Both comets appear inactive (i.e., lacking a coma) during the IRS observations (left column). \label{fig:spatial} }
\end{figure}

In Fig.~\ref{fig:spatial}, we plot the radial surface brightness profiles of all images of our target comets.  We include the MIPS image of Arend-Rigaux to demonstrate how emission from a dust tail affects the radial profile and point source fitting.  For comparison, we have included the spatial profile of asteroid 2004~BL$_{11}$ \citep{campins09-lowq} observed with the same IRS peak-up array as our comets (16-\micron{} array in peak-up acquisition mode).  For a comparison to the MIPS images, we have chosen asteroid (273) Atropos from the \spitzer{} archive (observed on 2008 Jan 20 at 1.02 au from the Sun).  These asteroids are important references as they provide real world spatial profiles of bright solar system point sources.  We reduced the asteroid data with the same procedures as used for the comet images and show their surface brightness profiles in Fig.~\ref{fig:spatial}.

Comparing the spatial profiles to the point-response functions (PRFs) of the instruments may reveal the presence of a coma or tail.  We use the term point-response function when we incorporate the effects of the detector and mosaicking process into a super-sampled image of the point-spread function \citep[cf.][]{makovoz05-extraction}.  Synthetic point-spread functions (PSFs) were generated using the \spitzer{} Science Center's STinyTim software\footnote{Available at the NASA/IPAC Infrared Science Archive, \url{http://irsa.ipac.caltech.edu/data/SPITZER/docs/dataanalysistools/}} at a one-tenth pixel scale.  The PSF changes significantly over the broad wavelength ranges of the IRS and MIPS filters.  To ensure the PSFs are appropriate for our targets, they were generated with template spectra based on each target's best-fit thermal model.  The NEATM models for Tempel~2 and Arend-Rigaux are described in Section~\ref{sec:spectra}, the Atropos model is from \citet{ryan10}, and the BL$_{11}$ model from \citet{campins09-lowq}.

The point-spread functions generated by STinyTim do not match the spatial profiles of our example asteroids BL$_{11}$ and Atropos.  This discrepancy exists because STinyTim does not account for any pixel-level instrument effects, or our image mosaicking procedures.   Smoothing the IRS PSF with a boxcar kernel of width 1.9~pixels produced a PRF that matched the spatial profile of asteroid BL$_{11}$.  We similarly smoothed the PSFs for each of the IRS peak-up images.\footnote{This smoothing function (1.9-pixel-wide boxcar) may only be valid for peak-up acquisition mode observations, i.e., not for peak-up imaging mode.}  The STinyTim MIPS PSF required a smoothing of 1.7~pixels to best approximate the Atropos profile; however, smoothing with a 1.3~pixel kernel better matched the Tempel~2 profile.  For consistency, we adopted 1.7~pixels for all MIPS observations. \citet{engelbracht07} used a similar sized filter (1.8 pixels) to match their MIPS 24-\micron{} observations of calibration star HD~159330.

Each observation's best-fit radial profile is included in Fig.~\ref{fig:spatial}, and the flux densities and effective radii given in Table~\ref{tab:fluxes}.  The IRS target flux densities have been color corrected by dividing by 0.98 (the pipeline is calibrated assuming a spectral energy distribution with $\nu F_\nu$ proportional to 1).  The magnitude of the color correction was computed using the best-fit NEATM models and the 16-\micron{}-filter transmission profile.  The photometry from the MIPS instrument, calibrated under different assumptions, do not require a color correction given our best-fit models.  Flux uncertainties include a 2\% absolute calibration uncertainty for IRS and 4\% for MIPS.

The comet cores and first diffraction rings are well matched by the smoothed synthetic PRFs confirming that they are dominated by point sources, with the Arend-Rigaux MIPS observation as an exception.  At this epoch, Arend-Rigaux  has a wide distribution of fluxes above its model PRF due to the presence of the dust tail at this epoch (Fig.~\ref{fig:ar-image}).  The Tempel~2 IRS profile exhibits the same deviations, but to a lesser degree due to the increased contrast between the nucleus and that comet's dust trail.  Finally, both the Arend-Riguax and Tempel~2 IRS profiles have deviations from the PRF beyond 6 pixels due to the gradient in the image background.  A gradient with the same morphology is seen in the red peak up arrays, which are offset from the target by about 1\arcmin, indicating it is not related to the comet.  We conclude there is no evidence for a dust coma or tail in the Tempel~2 or Arend-Rigaux IRS data sets based on this analysis.

{\renewcommand\baselinestretch{1}
  \begin{table}
    \small
    \caption{Best-fit near-Earth asteroid thermal model (NEATM) parameters.
      \label{tab:fluxes}}
    \begin{center}
      \begin{tabular}{@{\extracolsep{2mm}}lccccccc}
        \hline
        
        & \multicolumn{2}{c}{IRS}
        & \multicolumn{5}{c}{MIPS or IRS Peak-Up}
        \\\cline{2-3}\cline{4-8}

        Target
        & $\eta$
        & $\sigma_\eta$
        & $\lambda$
        & $F_\nu$
        & $\sigma_F$
        & $R$
        & $\sigma_R$ \\

        & 
        &
        & (\micron)
        & (mJy)
        & (mJy)
        & (km)
        & (km) \\
\hline
9P/Tempel 1            & 0.831                   & 0.005   & \nodata & \nodata & \nodata & \nodata & \nodata \\
10P/Tempel 2           & 0.813                   & 0.002   & 15.8 & 99.5                    & 2.0 & 4.48                    & 0.05 \\
                       &                         &         & 23.7 & 92.9                    & 3.8 & 4.35                    & 0.09 \\
49P/Arend-Rigaux       & 0.743                   & 0.005   & 15.8 & 27.4                    & 0.7 & 4.57                    & 0.06 \\
                       &                         &         & 23.7 & 1710\textsuperscript{b} &  68 & 5.40\textsuperscript{b} & 0.11 \\
(273) Atropos          & 0.97\textsuperscript{a} & 0.13    & 23.7 & 1296                    &  52 & 14.8                    & 0.3 \\
(276703) 2004 BL$_{11}$ & 2.45\textsuperscript{a} & 0.01    & 15.8 &  322                    &   6 & 0.60                    & 0.01 \\

\hline

      \end{tabular}
    \end{center}

\begin{footnotesize}
  Table columns|$\eta$, $\sigma_\eta$: NEATM IR beaming parameter
  based on spectra in this work (Tempel~1, Tempel 2, Arend-Rigaux), or
  based on the literature (Atropos, 2004 BL$_{11}$); $F_\nu$,
  $\sigma_F$: color-corrected best-fit flux density and uncertainty,
  from image photometry; $R$, $\sigma_R$, NEATM effective radius and
  uncertainty.

$^a$ The mean IR emissivity is 0.90 for these objects, to be
consistent with \citet{campins09-lowq} and \citet{ryan10}.

$^b$ This observation is contaminated by dust.

\end{footnotesize}

  \end{table}
}

\subsection{Spectroscopy}\label{sec:spectra}
\label{sec:spec-overview}
Given that each comet's spatial profile agrees with that of a point source, we fit the spectra of our targets with the NEATM, minimizing the $\chi^2$ statistic and allowing $\eta$ and the radius, $R$, to vary as free parameters.  The parameter uncertainties are based on the co-variance matrix returned by the least-squares routine \citep{bevington92}.  Our best-fit model spectra are presented in Fig.~\ref{fig:spec} and the best-fit $\eta$ values are given in Table~\ref{tab:fluxes}.  These $\eta$ values are used to convert the IRS peak-up photometry into effective radii, as discussed in Section~\ref{sec:imaging}.  Our value for Tempel~1, $\eta=0.860\pm0.005$, is equivalent to that derived by \citet{lisse05-nucleus}, $\eta=0.85$.

The Tempel~2 images have a prominent dust trail with a peak 24-\micron{} surface brightness of 0.6~\mjysr{} near the nucleus.  We use the 24-\micron{} image to estimate the contribution of the dust trail to our IRS spectrum.  We define an aperture using the IRS LL1 slit width (10.7\arcsec), our spectral extraction width (33\arcsec{} at 24~\micron{}, Section~\ref{sec:spec}), and orientation (long dimension 169\degr{} from the projected trail direction).  With this aperture, we measure the trail flux on both sides of the nucleus and average them together.  The trail signal is estimated to be $4.7\pm0.3$\% of the total spectral flux density at 24~\micron{} in the IRS spectrum.  We expect this fraction to increase with wavelength due to the wider spectral extraction aperture used at longer wavelengths (Section~\ref{sec:spec}).  The trail signal is $<$1\% of the total flux density in the SL slit due to the narrower slit width and extraction apertures.  Spherical trail dust, with radii near 0.1--1~mm, would have color temperatures near that of an isothermal sphere in local thermodynamic equilibrium.  The spectrum of the trail would thus have a cooler color temperature than the nucleus, which primarily re-radiates absorbed energy through the daytime hemisphere.  This spectral difference also increases the relative trail contribution at longer wavelengths.  The stronger dust contribution to the longer wavelengths introduces a bias in the IR beaming parameter of Tempel~2 toward larger values.  Comparing the Arend-Rigaux (dust free) and Tempel~2 (dust contaminated) beaming parameters suggests the bias is $<0.05$.

\begin{figure}
\centering
\includegraphics[width=0.9\textwidth]{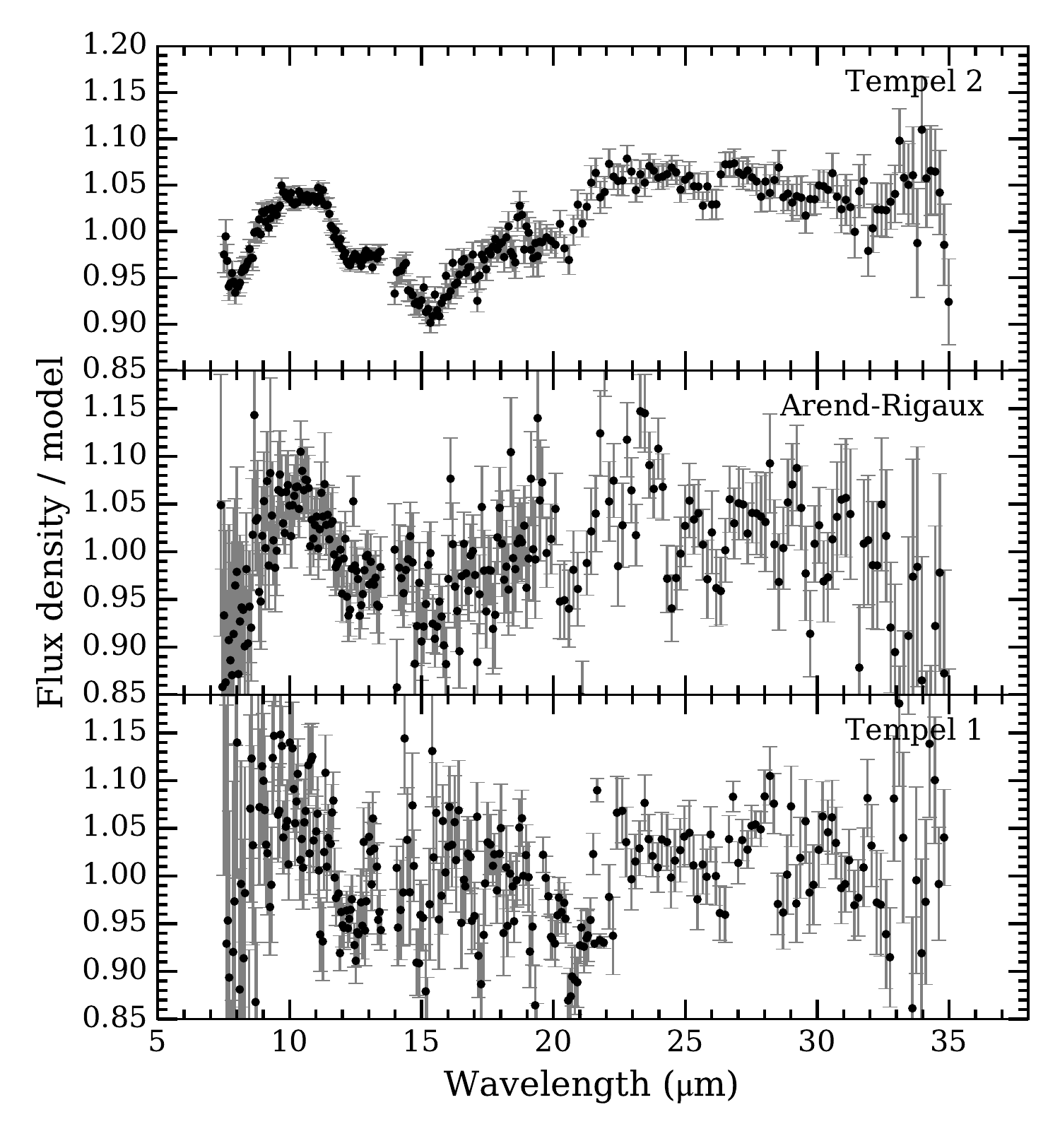}
\caption{\spitzer/IRS spectra of comets 10P/Tempel~2, 49P/Arend-Rigaux, and 9P/Tempel~1 normalized by their best-fit thermal emission models.  Due to their low signal-to-noise ratios, data points at $\lambda<7.4$~\micron{} have been removed from the Arend-Rigaux spectrum. \label{fig:emissivity}}
\end{figure}

Normalizing the comet spectra with their respective best-fit NEATM models reveals significant spectral features (Fig.~\ref{fig:emissivity}).    \citet{sivaraman15} examined Arend-Rigaux's 5--14~\micron{} spectrum based on the same data set.  They identified a curvature in the spectrum near 12--13~\micron{} and suggested this was an absorption feature due to amorphous water ice on the surface.  Rather than an absorption feature near 12--13~\micron, our examination of Tempel~2 and Arend-Rigaux together led us to identify the presence of a plateau from 9 to 11~\micron{} with a spectral contrast of about 10\%.  In addition, there is a minimum near 15~\micron{} with a depth of about 5\%, and local peaks near 19 and 23~\micron.  We interpret the 10-\micron{} plateau and the 19- and 23-\micron{} peaks as emission from silicate materials.  The 15-\micron{} feature remains unidentified and will be further discussed below.

\subsection{Comparison to photometry of comet nuclei}
\citet{fernandez13} observed 57 comet nuclei at 16 and 22~\micron{} with \spitzer's IRS peak-up arrays.  They fit their two-band photometry with the NEATM and derived a population averaged beaming parameter $\eta=1.03\pm0.11$.  In Section~\ref{sec:model}, we fit our nucleus spectra with the NEATM and found beaming parameters between 0.74 and 0.86 (Table~\ref{tab:fluxes}), 1.5 to 2.6$\sigma$ lower than the population average from \citeauthor{fernandez13}.  That all three spec\-tro\-scop\-i\-cal\-ly-derived beaming parameters are lower than the pho\-to\-met\-ri\-cal\-ly-derived beaming parameters suggests there is some systematic difference in the two approaches.  The best comparison would be through a comet-to-comet basis, but this is not possible as Tempel~1, Tempel~2, and Arend-Rigaux were not observed by \citeauthor{fernandez13}.  However, the analytical approaches are essentially the same: thermal emission data are fit using the NEATM assuming an emissivity of 0.95 and a fixed geometric albedo, allowing radius and beaming parameter to vary.  That \citeauthor{fernandez13} assumed a geometric albedo of 0.04 for all targets and that we used the known albedos of Tempel 1 and Tempel 2 is not significant; the difference in thermal emission from a surface with $A_p$=0.02 and one with $A_p$=0.04 is negligible.

To investigate the apparent discrepancy between the pho\-to\-met\-ri\-cal\-ly-derived and spec\-tro\-scop\-i\-cal\-ly-derived beaming parameters, we generated 16- and 22-\micron{} synthetic photometry based on the IRS spectra of Tempel~1, Tempel~2, and Arend-Rigaux using the IRS filter transmission profiles.  We then fitted a beaming parameter and radius to the photometry.  To make fits that are equivalent to the \citeauthor{fernandez13}\ nucleus survey, we assumed a fixed geometric albedo of 0.04 and a 3\% photometric uncertainty.  We derived $\eta=0.88\pm0.12$, $1.17\pm0.18$, and $0.97\pm0.13$ for Tempel~1, Tempel~2, and Arend-Rigaux, respectively.  Although the uncertainties are quite broad due to our artificial 3\% uncertainty factor, the photometrically retrieved $\eta$ values for Tempel~2 and Arend-Rigaux are higher than were obtained from the full spectra; the Tempel~1 $\eta$ value is essentially unchanged.  The photometry and models are presented in Fig.~\ref{fig:spec}.

Beaming parameters derived from either method may be affected by wavelength range and emissivity features.  To explore these effects, we performed additional tests with the NEATM and fit: (1) the short-wavelength orders only ($\lambda<14$~\micron); and (2) the long-wavelength orders only.  All derived $\eta$-values are presented in Table~\ref{tab:eta}.  There are two consistent trends among all the fits: the beaming parameters based on the shortest wavelengths are smaller than those that include longer wavelength data, and the fits that are only based on longer wavelengths have the highest values.  As one may expect, the most representative beaming parameters for NEATM modeling should be based on a wide range of wavelengths.  This is a partial explanation for the systematically higher beaming parameters derived from the synthetic 16- and 22-\micron{} photometry.

{\renewcommand\baselinestretch{1}
  \begin{table}
    \small
    \caption{Best-fit NEATM beaming parameters given different data sources.\label{tab:eta}}
    \begin{center}
      \begin{tabular}{l *{4}{cc}}
        \hline
        Target
        & \multicolumn{2}{c}{$\lambda<14$~\micron}
        & \multicolumn{2}{c}{All $\lambda$}
        & \multicolumn{2}{c}{$\lambda>14$~\micron}
        & \multicolumn{2}{c}{16 \& 22~\micron{}} \\
        
        & $\eta$ & $\sigma_\eta$
        & $\eta$ & $\sigma_\eta$
        & $\eta$ & $\sigma_\eta$
        & $\eta$ & $\sigma_\eta$ \\
\hline
Tempel 2     & 0.785 & 0.003 & 0.813 & 0.002 & 1.09 & 0.01 & 1.17 & 0.19 \\
Arend-Rigaux & 0.69  & 0.01  & 0.742 & 0.005 & 0.85 & 0.02 & 0.97 & 0.13 \\
Tempel 1     & 0.72  & 0.02  & 0.860 & 0.005 & 0.96 & 0.01 & 0.88 & 0.12 \\
\hline

      \end{tabular}
    \end{center}
  \end{table}
}

Aside from wavelength range, emissivity features provide an additional systematic offset.  As shown in Figs.~\ref{fig:emissivity} and \ref{fig:emissivity-trojans}, the 22-\micron{} region can have a significantly higher relative emissivity than the 16-\micron{} region, due to the presence of silicate materials.  We find that increasing the 16-\micron{} flux density by 1$\sigma$ (here, 3\%) decreases $\eta$ by 1$\sigma$ (here, $\sim$0.15), at least for our two-band photometry fits.  If the spectra in Fig.~5 are representative of the comet population, then a 3\% bias can easily be present in the \citet{fernandez13} photometry.  Lowering the \citeauthor{fernandez13} population mean beaming parameter by 1$\sigma$ from 1.03 to 0.92, results in a better agreement ($<$2$\sigma$) with each of our spectroscopically derived values.  Thus, we suggest emissivity features due to silicates are common in the comet population, and that the beaming parameters of \citet{fernandez13} are slightly biased by these features at the $\sim$1$\sigma$ level.  If true, then the effective radius estimates by \citeauthor{fernandez13}\ are higher than would otherwise be retrieved.  However, such an effect appears to be subtle, given the statistical agreement between the \citeauthor{fernandez13}\ survey and previous reflected light surveys of comet nuclei.

In contrast to this discussion, analysis of the nucleus of comet Schwass\-mann-Wach\-mann~1 by \citet{schambeau15}, using photometry spanning a broad wavelength range (5 to 70~\micron), results in a beaming parameter of 0.99$^{+0.26}_{-0.19}$.  The best-fit range is large, but it is centered near 1.0 and biased towards larger values.  Moreover, 12- and 22-\micron{} photometry of 22 nuclei with the \textit{WISE} spacecraft suggests $\eta$=1.0 to 1.2 \citep{bauer15}.  As an alternative explanation for the discrepancy between our results and those of \citeauthor{fernandez13}, \citeauthor{bauer15}, and \citeauthor{schambeau15}, the spectroscopy may be in error.  In Section~\ref{sec:spec}, we corrected apparent module-to-module offsets in our Tempel~2 and Arend-Rigaux spectra.  The corrections were derived by linear interpolation across the 14-\micron{} teardrop artifact present in the IRS spectra.  It is possible that the linear approximation is a poor one, and that the resulting scale factors are in error, which would have consequence on the best-fit NEATM models.  Fitting the unscaled Tempel~2 and Arend-Rigaux spectra does indeed raise the beaming parameters to near 0.9.  We do not have a solution for this aspect, but we surmise that spectral scaling is a better approach given potential uncertainties in the IRS pointing (Section~\ref{sec:spec}).  Future dedicated spectroscopic observations of these or other bare nuclei in the 5 to 30~\micron{} range with missions such as the \textit{James Webb Space Telescope} should be carefully obtained, and with supporting photometry if order-to-order corrections might be needed.

\subsection{Comparison to spectra of Jovian Trojan D-types, and other asteroids}

\begin{figure}
\centering
\includegraphics[width=0.95\textwidth]{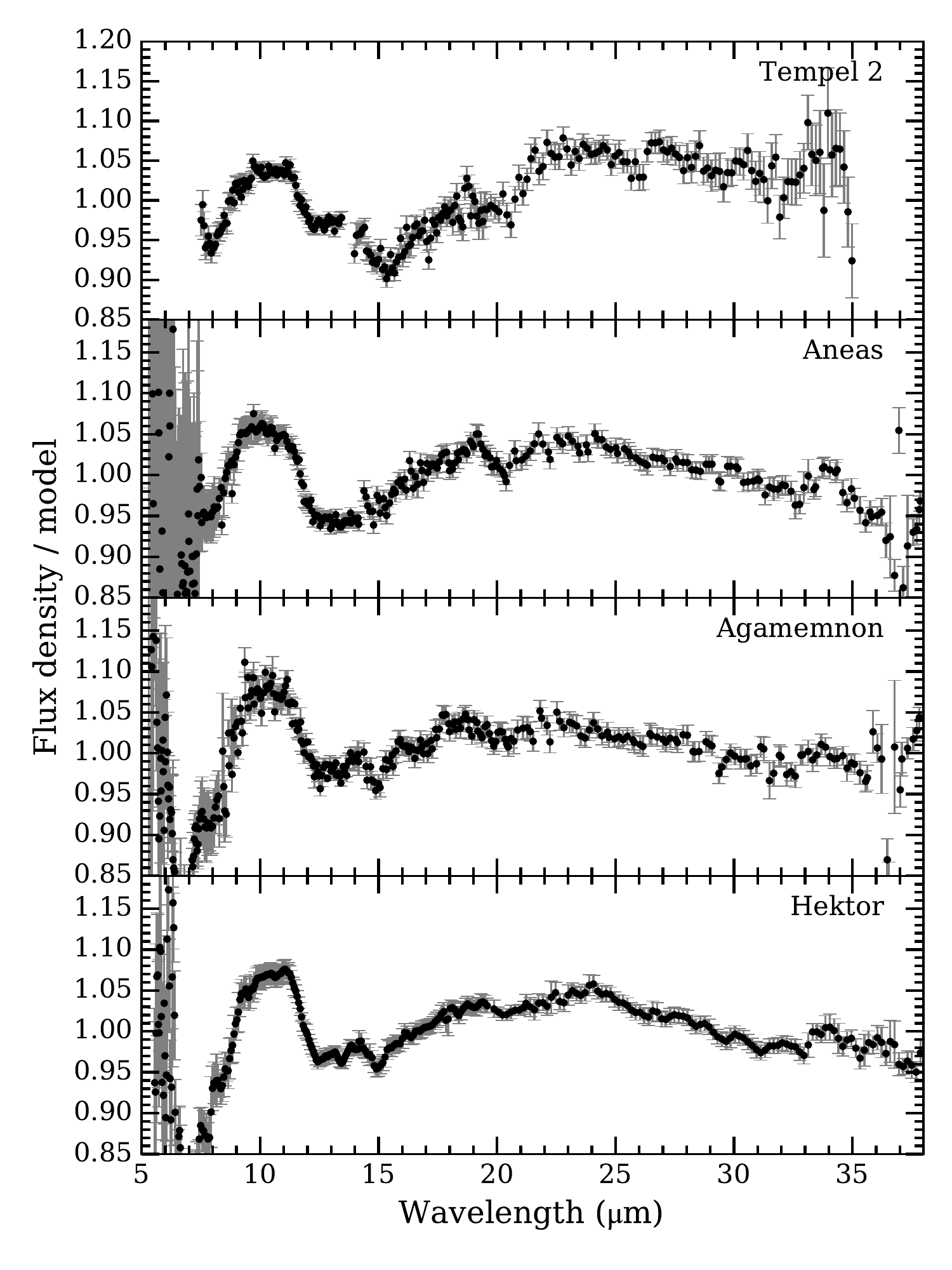}
\caption{\spitzer/IRS spectrum of the nucleus of comet 10P/Tempel~2, and of Jovian Trojans (911) Agamemnon, (624) Hektor, and (1172) Aneas normalized by their best-fit thermal emission models.  The asteroid spectra are from \citet{emery06}.  \label{fig:emissivity-trojans}}
\end{figure}

Comet nuclei tend to have D-type classifications based on their reflectance spectra (Section~\ref{sec:intro}).  We find that this similarity can be extended to their thermal emission spectra. Figure~\ref{fig:emissivity-trojans} shows our spectrum of Tempel~2 along with the spectra of Jovian Trojan D-types (624) Hektor, (911) Agamemnon, and (1172) Aneas from \citet{emery06}.  The 9- to 11-\micron{} plateau is slightly stronger in the Trojans.  Asteroid Hektor's feature has a trapezoidal shape like Tempel~2, but the Agamemnon and possibly the Aneas features are slightly rounded.  All four objects have the two emissivity peaks near 19 and 23~\micron.  The 15-\micron{} feature of Tempel~2 is the strongest, followed by those of Hektor and Agamemnon.  Asteroid Aneas's 15-\micron{} feature is weak, if present at all.  We summarize the parameters of the emissivity features of our three comets and the three Trojan D-types in Table~\ref{tab:features}.

Asteroid Hektor and comet Tempel~2 have the most similar spectra.  The dust grains in comet nuclei are primarily composed of silicates of mixed Mg and Fe content, Mg-rich silicates, carbonaceous dust, and metal sulfides \citep{wooden02, hanner10, frank14}. The spectral similarity between Hektor and Tempel~2 suggests that at least some Jovian Trojan D-types have surface compositions and structures similar to comet nuclei.

{\renewcommand\baselinestretch{1}
  \begin{table}
    \small
    \caption{Summary of spectral features in comet nuclei and
      comparison to select asteroids.\label{tab:features}}
    \begin{center}
      \makebox[\textwidth]{
      \begin{tabularx}{1.2\textwidth}{lcccccc}
        \hline
        Target
        & 9--11~\micron
        & 15-\micron
        & 20-\micron
        & 26-\micron
        & 28-\micron
        & 34-\micron \\

        & plateau
        & minimum
        & minimum
        & minimum
        & maximum
        & maximum\textsuperscript{a} \\
\hline

Tempel 2     & trapezoidal    & present & present & strong & none   & ? \\
Arend-Rigaux & present        & present & present & ?      & ?      & none \\
Tempel 1     & present        &   ?  & ?       & ?      & ?      & ? \\
Hektor       & trapezoidal    & present & present & none   & subtle & present \\
Agamemnon    & rounded        & present & present & none   & none   & ? \\
Aneas        & rounded        & none & present & none   & none   & present \\

\hline

      \end{tabularx}}
    \end{center}
    \begin{footnotesize}
      \textsuperscript{a}The apparent presence of the 34~\micron{}
      peak, being at the edge of the wavelength range, is strongly
      affected by the thermal model in our comet spectra.
    \end{footnotesize}
  \end{table}
}

Spectral features exist in mid-infrared spectra of many asteroids.  We examined the \spitzer{} IRS spectra of: E-type asteroid (2867) \v{S}teins \citep{barucci08, groussin11}; M-type asteroid (21) Lutetia \citep{barucci08}; the 28 C-, X-, V-, and S-type spectra of \citet{marchis12}; the X-type asteroid (617) Patroclus \citep{mueller10}; the T-type (308) Polyxo \citep{dotto04}; and the V-type (956) Elisa \citep{lim11} to compare with our comet nuclei spectra.  Of these asteroids, the C- and X-type spectra presented by \citet{marchis12} are most similar to Tempel~2, with broad maxima centered at 10~\micron{} and 16--17~\micron{}.  However, these peaks tend to be more rounded in the asteroid spectra.  \citet{mueller10} already noted similarities between X-type Patroclus and D-type Hektor, the former asteroid having weaker features.  Except for the Jovian Trojans Hektor, Agamemnon, and Patroclus \citep{emery06, mueller10}, none of the above investigations showed any 15-\micron{} feature similar to that seen in the spectra of Tempel~2 and Arend-Rigaux.

\subsection{Comparison to spectra of comet comae}\label{sec:comae}

\citet{emery06} noted striking similarities between their D-type Jovian Trojan spectra and the spectra of the comae of comets Hale-Bopp and Schwass\-mann-Wach\-mann~1 \citep{crovisier97-science, stansberry04}, and hypothesized that the surfaces of the D-types are comprised of fine-grained silicate material embedded in a transparent or high porosity surface.  To better understand this similarity, we compare the spectrum of the surface of Tempel~2 to the spectra of comet comae.

In addition to Hale-Bopp and Schwass\-mann-Wach\-mann~1, we selected comets from the literature that were observed with \spitzer{}/IRS at high signal-to-noise ratios: Tempel~1 \citep[pre-\di{} experiment, nucleus removed;][]{lisse06, kelley09-jfc}, 17P/Holmes \citep[post-mega-outburst;][]{reach10}, 73P/Schwass\-mann-Wach\-mann~3 fragments B and C \citep{sitko11}, and Oort cloud comet C/2007 N3 (Lulin) \citep{woodward11}.  The Schwass\-mann-Wach\-mann~1 spectrum is based on our own processing and spectral extraction, following the methods of \citet{kelley06-comets}.  This includes the removal of a model nucleus, based on the nucleus parameters of \citet{schambeau15}.  The properties of the spectra of the comet comae are summarized in Table~\ref{tab:comae}.

\subsubsection{Apparent continuum temperature}\label{sec:continuum}

\begin{figure}
\centering
\includegraphics[width=0.9\textwidth]{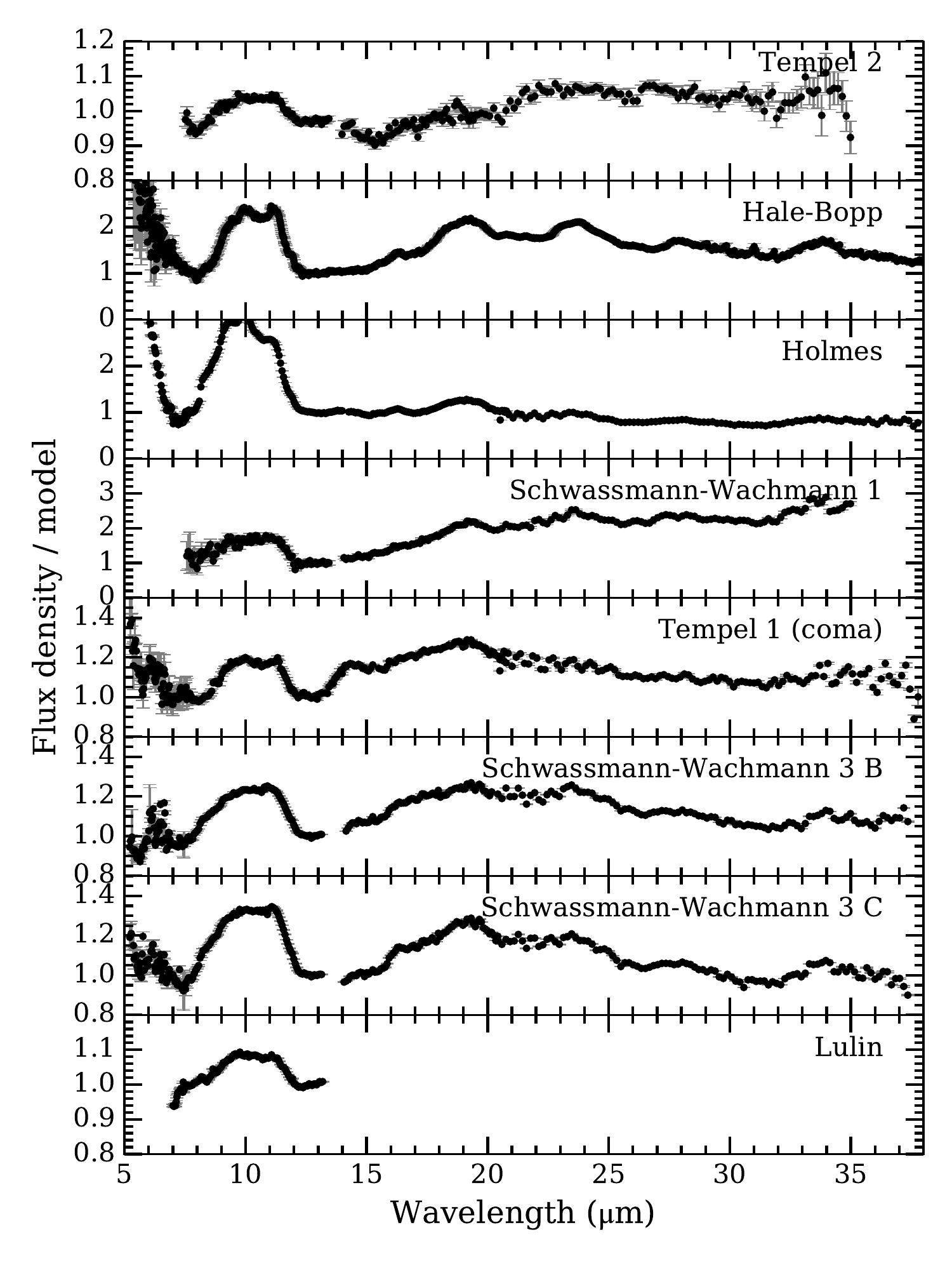}
\caption{Spectra of the nucleus of comet 10P/Tempel~2, and of the comae of comets 9P/Tempel~1, 17P/Holmes (post-mega-outburst), 29P/Schwass\-mann-Wach\-mann~1, 73P/Schwass\-mann-Wach\-mann~3 fragments B and C, C/1995 O1 (Hale-Bopp) and C/2007 N3 (Lulin).  See Table~\ref{tab:comae} and the text for data sources.  The comatic spectra have been normalized by a scaled Planck function fit to the data points near 8 and 13~\micron.  Note the varying y-axis limits between each subplot. \label{fig:emissivity-comae}}
\end{figure}

Each coma spectrum is fitted with a scaled Planck function, a common practice in comet infrared spectral analysis \citep[e.g.,][]{bregman87, hanner94-silicates, sitko04}.  We used the 7.5--8.0 and 12.5--13.0-\micron{} regions to establish the local 10-\micron{} pseudo-continuum.  We also fitted and include in Table~\ref{tab:planck} the spectra of the comet and asteroid surfaces to facilitate a one-to-one comparison.  The best-fit temperatures are given in Table~\ref{tab:planck} and each comet's spectrum is normalized by the best-fit Planck function and plotted in Fig.~\ref{fig:emissivity-comae}.  Column 4 in Table~\ref{tab:planck} lists the color-temperature excess $T_c/T_{BB}$, i.e., the best-fit color temperature, $T_c$, divided by the temperature of a blackbody sphere in local thermodynamic equilibrium with insolation $T_{BB}=278~{\mathrm K}/\sqrt{r_h}$, where $r_h$ is in au \citep{gehrz92}.  We have ignored the best-fit temperature for comet Schwass\-mann-Wach\-mann~1 as it does not agree with the longer wavelengths ($\lambda>15$~\micron).  The 10-\micron{} spectrum is strongly dominated by the nucleus \citep{stansberry04, schambeau15}, and its high apparent temperature is likely due to uncertainties in the nucleus subtraction.

There is a clear dichotomy in our target sample: the spectra of the asteroid and comet nucleus surfaces have $T_c/T_{BB}$ values of 1.35--1.41, yet the comae are cooler at 1.12--1.29.  This grouping is expected.  Slowly-rotating surfaces with low thermal inertia re-radiate most of their absorbed sunlight on the daytime hemisphere nominally raising the color temperature by a factor of $2^{1/4}$ (1.19), whereas the dust grains in comet comae, with numerous radii less than 100~\micron{} in size, are more isothermal and can re-radiate in all directions.  The similarity in the normalized temperatures between Tempel~2, Arend-Rigaux, and Hektor gives further evidence that these spectra are dominated by nucleus emission, rather than a spatially unresolved coma.  A high color-temperature for comet Arend-Rigaux was already measured by \citet{tokunaga85}, and they also concluded they observed nucleus-dominated photometry.

%%%%%%%%%%%%%%%%%%%%%%%%%%%%%%%%%%%%%%%%%%%%%%%%%%%%%%%%%%%%%%%%%%%%%%%%%%%%%%%%
% Comet comae for comparison
%%%%%%%%%%%%%%%%%%%%%%%%%%%%%%%%%%%%%%%%%%%%%%%%%%%%%%%%%%%%%%%%%%%%%%%%%%%%%%%%
{\renewcommand\baselinestretch{1}
  \begin{table}
    \footnotesize
    \caption{Comet comae spectra\textsuperscript{a} selected for comparison to nucleus spectra.\label{tab:comae}}
    \begin{center}
      \makebox[\textwidth]{
      \begin{tabularx}{1.2\textwidth}{@{\extracolsep{0pt}}lcccp{2in}}
        \hline
        Target
        & Class
        & Instrument
        & $r_h$
        & Note \\

        &
        &
        & (au)
        \\
\hline

9P/Tempel 1 & JF & \spitzer/IRS & 1.5 & Pre-\di{} excavation, model nucleus removed. \\
17P/Holmes & JF & \spitzer/IRS & 2.5 & Post-mega-outburst.\\
29P/Schwassmann-Wachmann 1 & Centaur & \spitzer/IRS & 5.7 & Model nucleus removed. \\
73P/Schwassmann-Wachmann 3-B & JF & \spitzer/IRS & 1.5 & 16 days post-outburst. \\
73P/Schwassmann-Wachmann 3-C & JF & \spitzer/IRS & 1.5 & \\
C/1995 O1 (Hale-Bopp) & OC & \iso/SWS & 2.8 \\
C/2007 N3 (Lulin) & OC & \spitzer/IRS & 1.9 \\

\hline

      \end{tabularx}}
    \end{center}

    \begin{footnotesize}
      Table columns|Class: dynamical classification (JF) Jupiter-family comet
      centaur, or (OC) Oort cloud comet; $r_h$: heliocentric distance.

      \textsuperscript{a} See text for data sources.

    \end{footnotesize}

\end{table}
}

\subsubsection{Emissivity features}\label{sec:silicatefeature}
\begin{figure}
\centering
\includegraphics[width=0.9\textwidth]{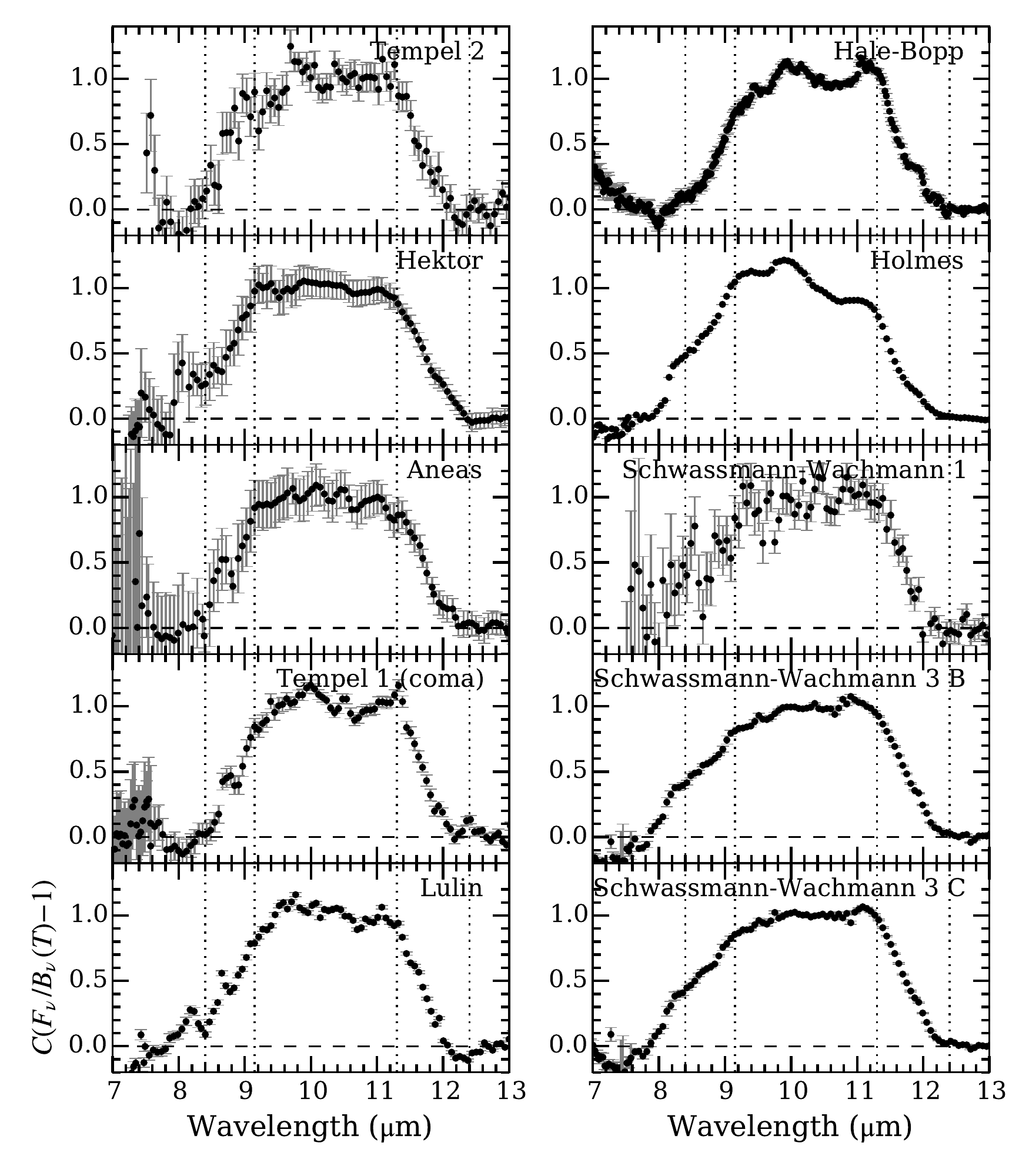}
\caption{The 10-\micron{} plateau from the nucleus of comet 10P/Tempel~2, asteroid (624) Hektor, the (pre-Deep Impact) coma of comet Tempel~1, and the comae of comets C/2007~N3 (Lulin), C/1995~O1 (Hale-Bopp), 17P/Holmes, 29P/Schwass\-mann-Wach\-mann~1, and fragments B and C of 73P/Schwass\-mann-Wach\-mann~3.  Each spectrum has been offset and normalized to span the range 0 to 1.0 as described in Section~\ref{sec:silicatefeature}.  Vertical dotted lines are positioned to mark the edges of features in the Hektor spectrum. \label{fig:bbnorm}}
\end{figure}

The 10-\micron{} region has the strongest feature observed at the best signal-to-noise ratio.  Figure \ref{fig:bbnorm} presents the 7--13~\micron{} spectra of the surface of comet Tempel~2, asteroids Hektor and Aneas, and the comae of our comparison comets.  All spectra have been normalized by the Planck function fit in Section~\ref{sec:continuum}, then offset and scaled to span the range 0 to 1.0.  The comae of comets Holmes, and Hale-Bopp, and to a lesser extent Tempel~1, have pronounced emission from Mg-rich silicates at 10.0 and 11.2~\micron{}.  There is a peak near 9.8~\micron{} in the Tempel~2 spectrum, but no 11.2~\micron{} peak.   On broader wavelength scales, the 10-\micron{} plateaus appear to fall into two groups: (1) trapezoidal and similar to the feature in the spectrum of the surface of Tempel~2 (Hektor, Aneas, comets Tempel~1, Holmes, and Hale-Bopp), or (2) the same but with a short-wavelength shoulder that is more rounded due to enhanced emission at 8 to 8.5~\micron{} (Schwass\-mann-Wach\-mann~1, Schwass\-mann-Wach\-mann 3-B, 3-C, and Lulin).

We examined the mineral decompositions of the spectra of comets Holmes, Hale-Bopp, Schwass\-mann-Wach\-mann~1, Schwass\-mann-Wach\-mann~3, and Lulin \citep{harker02-hb, reach10, sitko11, woodward11, schambeau15} but found no clear trend that would explain the differences.  The analysis of the 7--13-\micron{} spectra of comet Hale-Bopp by \citet{wooden99} indicates the shape of the short-wavelength shoulder is controlled by the relative amounts of amorphous and crystalline pyroxene (see their Figs.~5, 6, and 7), but in other comets the coma grain size distribution must also play a role.

In Table~\ref{tab:planck} we summarize the 10-\micron{} feature of each target with the wavelength of onset (the short-wavelength edge of the whole feature), crest (the short-wavelength edge of the plateau), cliff (the long-wavelength edge of the plateau), and terminus (the long-wavelength edge of the whole feature).  The positions were identified by visual inspection.  Each spectrum's crest, cliff, and terminus occurs within 0.1~\micron{} of those of the other targets.  As already suggested in Section~\ref{sec:comae}, such differences may be the result of composition, or grain size and packing variations, the latter of which can change the origin of photons with wavelengths near the transition between surface and volume scattering domains \citep{hunt68, salisbury92}.  The largest variation occurs at the onset, ranging from 7.8 (Schwass\-mann-Wach\-mann~3 B and C) to 8.4~\micron{} (Aneas).  The surfaces, however, all have similar onsets at 8.2 to 8.4~\micron.  The best match for Tempel~2 is Aneas and the coma of Tempel~1.

Finally for the 10-\micron{} region, we identify a subtle narrow peak at 12.4 to 12.5~\micron{} in the Tempel~2 surface and the Tempel~1 coma spectra.  A similar feature was seen in Gemini/Michelle spectra of material excavated from Tempel~1, but remains unidentified \citep{harker07}.

{\renewcommand\baselinestretch{1}
  \begin{table}
    \small
    \caption{Summary of the 10-\micron{} features.\textsuperscript{a}\label{tab:planck}}
    \begin{center}\makebox[\textwidth]{
      \begin{tabularx}{1.3\textwidth}{lcccccccc}
        \hline
        Target
        & Type
        & $T_c$
        & $\sigma_T$
        & $T_c/T_{BB}$
        & Onset
        & Crest
        & Cliff
        & Terminus \\

        &
        & (K)
        & (K)
        &
        & (\micron)
        & (\micron)
        & (\micron)
        & (\micron) \\
\hline

Arend-Rigaux             & nucleus  & 208.4 & 2.2 & 1.41 & \nodata & \nodata & (11.3) & (12.2) \\
Tempel 2                 & nucleus  & 241.8 & 0.4 & 1.39 & (8.2)   & (9.0)   & 11.3   & 12.2   \\
Agamemnon                & asteroid & 162.6 & 0.4 & 1.38 &         &         &        &        \\
Aneas                    & asteroid & 161.1 & 0.3 & 1.39 & 8.4     & 9.2     & 11.4   & 12.2   \\
Hektor                   & asteroid & 164.3 & 0.3 & 1.35 & (8.4)   & 9.2     & 11.3   & 12.4   \\
Hale-Bopp                & coma     & 214.5 & 0.5 & 1.29 & 8.2     & 9.2     & 11.2   & 12.3   \\
Holmes                   & coma     & 204.0 & 0.0 & 1.16 & 8.0     & 9.2     & 11.2   & 12.2   \\
Lulin                    & coma     & 226.7 & 0.1 & 1.12 & 8.0     & (9.6)   & 11.3   & 12.2   \\
Schwassmann-Wachmann 1   & coma     & 198.0\textsuperscript{b} & 4.3 & 1.70\textsuperscript{b}
                                                         & (8.0)   & (9.3)   & 11.4   & 12.2   \\
Schwassmann-Wachmann 3 B & coma     & 301.9 & 0.2 & 1.18 & 7.8     & 9.1     & 11.2   & 12.2   \\
Schwassmann-Wachmann 3 C & coma     & 268.0 & 0.2 & 1.17 & 7.8     & 9.1     & 11.2   & 12.2   \\
Tempel 1                 & coma     & 273.5 & 0.7 & 1.39 & 8.2     & 9.2     & 11.3   & 12.2   \\
\hline

      \end{tabularx}}
    \end{center}
    \begin{footnotesize}
      Table columns | $T_c$, $\sigma_T$: best-fit color temperature
      and uncertainty; $T_c/T_{BB}$: color-temperature excess, the
      ratio of the color temperature to the temperature of an
      isothermal spherical blackbody at the same heliocentric
      distance; Onset: the short-wavelength edge of the silicate
      feature; Crest: the short-wavelength edge of the plateau; Cliff:
      the long-wavelength edge of the plateau; Terminus: the
      long-wavelength edge of the silicate feature.

      \textsuperscript{a} Values in parentheses are estimates that are
      affected by low data quality or unclear spectral structure.

      \textsuperscript{b} Although this is the best-fit temperature
      for this comet, we do not consider it to be robust, as evidenced by its
      poor accounting for the longer wavelength emission in
      Fig.~\ref{fig:emissivity-comae}.

    \end{footnotesize}
  \end{table}
}

At longer wavelengths, the emission features from Mg-rich silicate grains at 16, 19, 23, 28, and 34~\micron{} are stronger in the comae spectra than in the surface spectra.  For example, the 19~\micron{} peak has relative strengths of 10--25\% in most of the comae, but only $\sim$5\% at Tempel~2.  In general, this may indicate a grain size difference---the relative strengths of crystalline silicate resonances vary with grain size \citep{lindsay13-paper1}---or an abundance difference.  The Tempel~2 features are not well aligned with the coma features, e.g., we estimate the Tempel~2 emissivity peaks are near 16.5, 18.9, 21.7--23.0, and 26.5~\micron{}, whereas the corresponding Hale-Bopp peaks are near 16.2, 19.1, 23.7, and 27.8~\micron.  Such shifts may be the effect of grain-grain contact on the surface of the object, causing light scattering between grains.

\subsection{The 15-\micron{} feature}\label{sec:fifteen}
\begin{figure}
\centering
\includegraphics[width=0.75\textwidth]{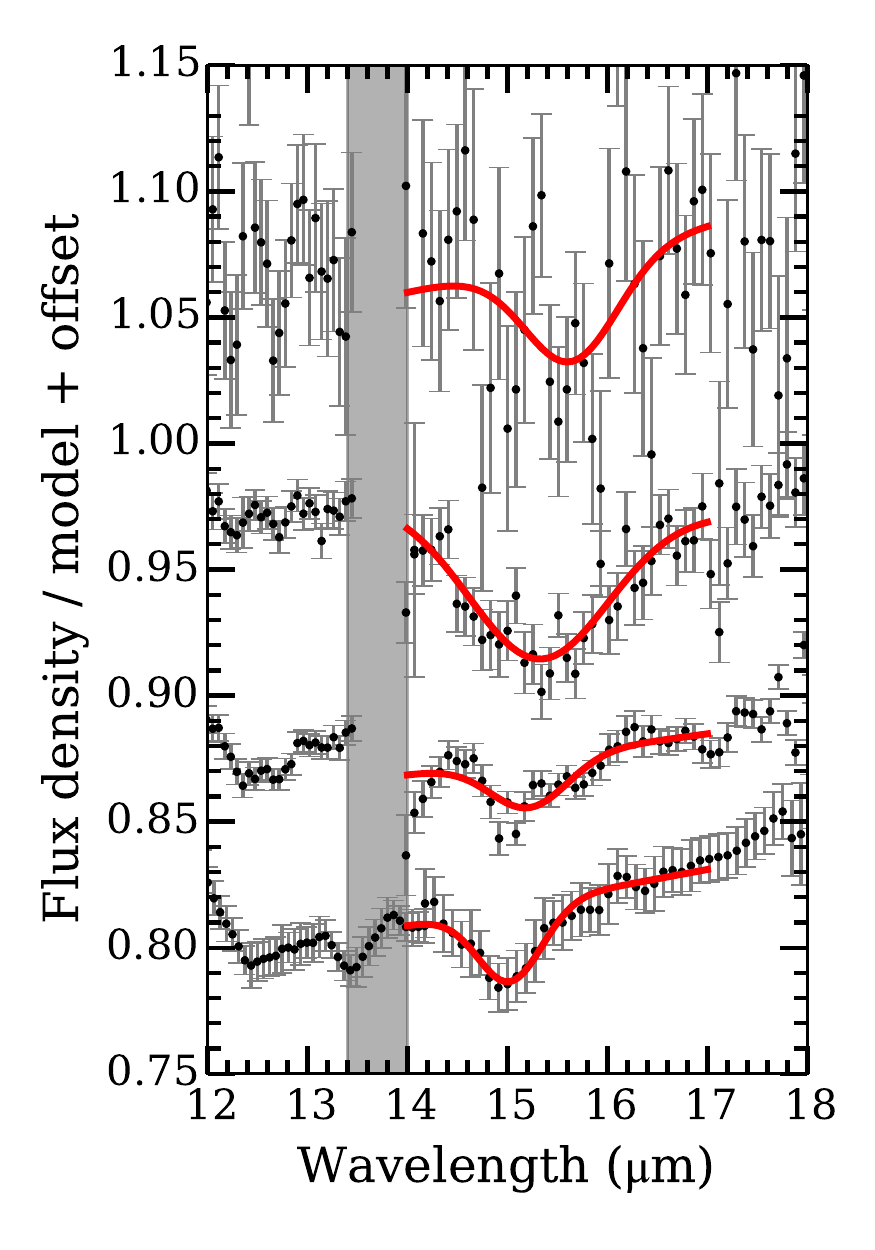}
\caption{The 12--18~\micron{} region for (top to bottom) 49P/Arend-Rigaux, 10P/Tempel~2, our reduction of (624) Hektor, and Hektor from \citet{emery06}.  The spectra have been offset for clarity.  Solid lines depict each spectrum's best-fit Gaussian function plus linear continuum.  The gray region marks the IRS teardrop artifact location (Section~\ref{sec:spec}).  \label{fig:fifteen}}
\end{figure}

A 15~\micron{} emissivity minimum is shown in detail in Fig.~\ref{fig:fifteen}.  We estimate the approximate wavelength position of the feature by fitting a Gaussian function plus a linear continuum to each 14--17-\micron{} spectrum.  The best-fit parameters and equivalent widths are presented in Table~\ref{tab:fifteen}.  The locations of Arend-Rigaux and Tempel~2 minima are consistent with each other.

{\renewcommand\baselinestretch{1}
  \begin{table}
    \small
    \caption{Best-fit 15-\micron{} Gaussian function amplitudes, positions, and feature equivalent widths.\label{tab:fifteen}}
    \begin{center}
      \begin{tabular}{lcccccc}
        \hline
        Target
        & $A$
        & $\sigma_A$
        & $\mu$
        & $\sigma_\mu$
        & $EW$
        & $\sigma_{EW}$
        \\
        % target
        & % A
        & 
        & ($\micron$) % mu
        & ($\micron$)
        & ($\micron$) % EW
        & ($\micron$) % EW
        \\
\hline

Tempel 2                       & -0.105 & 0.057 & 15.31 & 0.12 & -0.101 & 0.004 \\
Arend-Rigaux                   & -0.049 & 0.043 & 15.64 & 0.24 & -0.046 & 0.011 \\
Hektor, S12\textsuperscript{b} & -0.024 & 0.006 & 15.03 & 0.07 & -0.023 & 0.003 \\
Hektor, S18\textsuperscript{b} & -0.019 & 0.005 & 15.22 & 0.06 & -0.018 & 0.002 \\

\hline

      \end{tabular}
    \end{center}
    \begin{footnotesize}
      \textsuperscript{a} Column headings: Gaussian function amplitude and uncertainty ($A$, $\sigma_A$), central wavelength and uncertainty ($\mu$, $\sigma_\mu$), and feature equivalent width and uncertainty ($EW$, $\sigma_{EW}$).

      \textsuperscript{b} The ``Hektor, S12'' spectrum \citep{emery06}
      was processed with version S12 of the IRS reduction pipeline.
      The ``Hektor, S18'' spectrum (this work) was processed with
      version S18.

    \end{footnotesize}
  \end{table}
}

There have been several updates to the IRS data reduction pipeline since the study of Trojan asteroids by \citet{emery06}, which used version S12.0.  To investigate whether the spectrum of the 15-\micron{} feature in Hektor is affected by the improved calibrations, we downloaded the Hektor observations processed with the latest IRS pipeline (S18.18, the same as Tempel~2 and Arend-Rigaux) from the \spitzer{} archive.  The spectra were reduced as outlined in Section~\ref{sec:obs}.  The 15-\micron{} feature was still present, although the shape and central wavelength changed from $15.03\pm0.07$ to $15.22\pm0.06$.  The position of the feature in the updated spectrum is consistent with the minimum in the spectrum of Tempel~2.  The apparent feature at 13.4~\micron{} in the original \citet{emery06} spectrum (bottom of Fig.~\ref{fig:fifteen}) is likely the IRS teardrop artifact (Section~\ref{sec:spec}).

Some \spitzer/IRS observations of young stellar objects show \coo{}-ice absorption near 15~\micron{} \citep[e.g.,][]{bergin05, quanz07}.  The \coo{}-ice feature has a minimum at 15.2~\micron, consistent with the spectra of both Hektor and Tempel~2.  However, the surfaces of our objects are too warm for \coo{} ice to be present in any significant abundance.  \coo{} readily sublimates at $T>80~K$, and to see this feature in absorption it must be obscuring a source of warm continuum.  \coo{} could be replenished from a comet's interior, but this does not account for the presence of \coo{} ice on the Trojan asteroids.  If the volatile \coo{} is present, then the less volatile water ice should also be present.  Initial studies by \citet{emery03} and \citet{yang07} found little evidence for water ice on their surfaces. More recently, \citet{brown16} has demonstrated that the less red Trojan asteroids have a wide 3-\micron{} absorption feature in their reflectance spectra, but he argues that water ice is unlikely in such a substantial fraction of the population and suggests this feature is due to N--H bonds rather than water ice.  Finally, surface \coo{} ice, whether at 5~au or closer to the Sun, should induce cometary activity, but this has not been reported for any Jovian Trojan asteroid.

\begin{figure}
\centering
\includegraphics[width=0.5\textwidth]{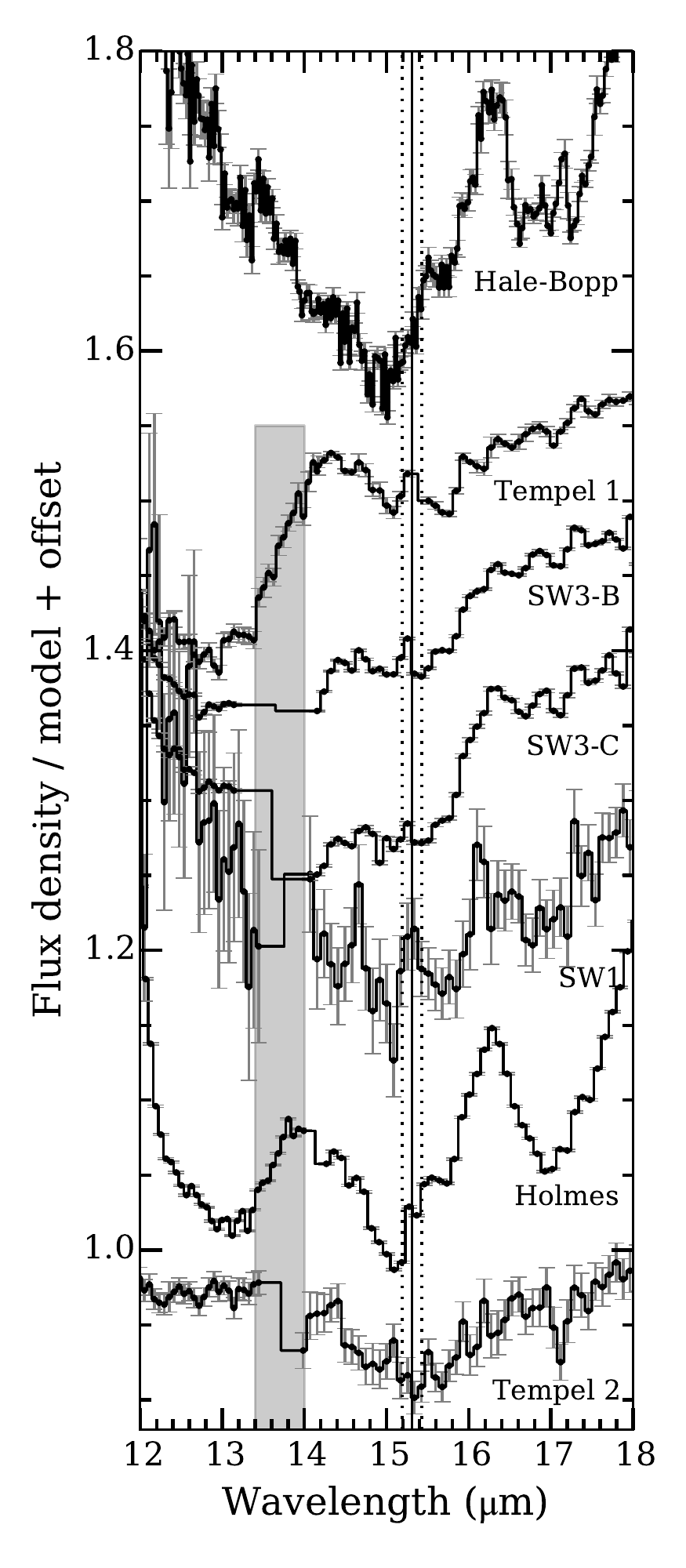}
\caption{The 12--18~\micron{} region from the nucleus of comet 10P/Tempel~2, and from comet comae: C/1995~O1 (Hale-Bopp), 9P/Tempel~1 (pre-\textit{Deep Impact} encounter), 73P/Schwass\-mann-Wach\-mann~3 fragments B and C, 29P/Schwass\-mann-Wach\-mann~1, and 17P/Holmes.  The spectra have been offset for clarity.  Vertical lines mark the minimum and 1-$\sigma$ uncertainty of the Tempel~2 feature.  See Fig.~\ref{fig:fifteen} for details and references.  The gray region marks the IRS teardrop artifact location in \textit{Spitzer}/IRS spectra (Section~\ref{sec:spec}).  \label{fig:fifteen-comets}}
\end{figure}

A 15-\micron{} feature has not been previously described in any spectrum of a cometary coma.  To investigate, we present spectra of the above comet comae in Fig.~\ref{fig:fifteen-comets}.  The spectrum of the nucleus of Tempel~2 was normalized by its NEATM model, and the comet comae spectra were normalized by a Planck function simultaneously fit to the wavelength regions $<$8, 12.5--16, and 21--23~\micron.  Except for Tempel~2, there are no clear 15-\micron{} minima, with the caveat that the broad 16.2-\micron{} emission from Mg-rich olivine in several comets may be obscuring their respective 15-\micron{} features.  Comet Tempel~1 has the only spectrum with a possible broad 15.3-\micron{} minimum, although this interpretation is not straightforward due to a narrow local maximum at the same location.

\section{Summary}\label{sec:summary}

We presented spectra and images of comets 10P/Tempel~2 and 49P/Arend-Rigaux, observed with the \spitzer/IRS instrument.  The spectra of these comets have high 10-\micron{} color-temperature excesses, higher than those seen for comet comae, but similar to those of asteroid surfaces.  The contemporaneously obtained mid-infrared images of these comets have radial surface brightness profiles that agree with those expected from point sources.  Their measured mid-infrared flux densities are commensurate with object radii similar to those already published in the literature using other techniques.  We conclude that the IRS spectra of these two comets represent the thermal emission of bare nuclei, i.e., free from coma contamination.  We estimated, however, that the spectrum of comet Tempel~2 does has a small ($\sim$5\%) contribution from the comet's dust trail at the longest wavelengths in our spectrum.r

Including previous spectra of the nucleus of comet 9P/Tempel~1 \citep{lisse05-nucleus}, we fit each spectrum with a near-Earth asteroid thermal model (NEATM) to derive their effective surface beaming parameters.  We found a systematic discrepancy between our spec\-tro\-scop\-i\-cal\-ly-derived IR beaming parameters and those derived for the JFC population by \citet{fernandez13} based on 16- and 22-\micron{} photometry.  The spec\-tro\-scop\-i\-cal\-ly-derived beaming parameters are near 0.8, meanwhile the \citet{fernandez13} study found a population mean of 1.03$\pm$0.11.  We suggest this may be due to the limited wavelength range of the two-band photometry, and the presence of emissivity features throughout the 7--30~\micron{} range.  The \citet{fernandez13} derived radii are consistent with optical surveys of comet nuclei, so any bias caused by the potentially high beaming parameter is statistically unimportant to the nucleus sizes.  However, we necessarily applied scale factors to our spectra to correct for apparent IRS module-to-module offsets.  This approach, based on a linear interpolation over a known instrument artifact, may also be in error.  New observations of bare nuclei by the MIRI instrument on the \textit{James Webb Space Telescope} \citep{rieke15} should help us understand this discrepancy.

By normalizing our spectra with a best-fit thermal model, we find emissivity features in which we identify a 9--11~\micron{} trapezoidal plateau, and weaker peaks near 18--19, 22--24, and 34~\micron{}.  These are the signatures of silicates, including Mg-rich olivine.  That all three comets inspected, Tempel~1, Tempel~2, and Arend-Rigaux, exhibit a 10-\micron{} plateau indicates that its presence is not correlated with the unusually low activities of the larger nuclei (Tempel~2 and Arend-Rigaux).  This interpretation is in contrast to that of \citet{sivaraman15}, who conclude a broad depression near 13~\micron{} in the 5--14-\micron{} spectrum of Arend-Rigaux indicates the presence of amorphous water ice.

There are strong similarities between the emissivity spectra of our comet nuclei and those of D-type asteroids.  \citet{emery06} has already compared three D-type Jovian Trojan asteroids to comet comae and found the same similarities.  That asteroid surfaces and dust in comet comae present the same features in emission is not completely expected, even if they have the same composition.  \citet{emery06} consider this point in detail, and conclude that the Trojan surfaces are either highly porous, or have fine-grained silicate species that are suspended in a mid-infrared-transparent matrix.  \citet{vernazza12} adds that the low surface thermal inertia and red reflectance spectra of these asteroids provides additional evidence for highly porous surfaces.  We note that the bulk densities of comets are also low suggesting porosities near 80\% \citep{richardson07, ernst07, thomas13-tempel1, thomas13-hartley2, sierks15}, and the mid-infrared similarities between the Trojan and comet surfaces further emphasize that high porosity is an important characteristic.

We compared \spitzer{} and \iso{} spectra of comet comae to those of our comet surfaces.  We find that the 10-\micron{} silicate features of comet comae can be described as either having a trapezoidal shape, or having a more rounded or triangular shape.  The latter is apparently due to enhanced emission near 8 to 8.5~\micron.  There has been evidence for a variety of shapes in previous ground-based coma spectroscopy (cf. \citealt{hanner94-silicates} and \citealt{sitko04}), but none seem to have the more rounded shapes of Schwass\-mann-Wach\-mann~1 or Schwass\-mann-Wach\-mann~3.  We will examine this aspect of comet comae further in a future paper.  The 10-\micron{} silicate features of our comet surfaces and those of Jovian Trojans Hektor and Aneas better match the comae with more trapezoidal-shaped silicate bands.

An emissivity minimum near 15.2~\micron{} (2--5\% depth) is seen in the spectra of Arend-Rigaux and Tempel~2.  The feature is also found in the spectra of Jovian Trojans Hektor, Agamemnon (D-type), and Patroclus (X-type), but we did not find it in the spectra of other asteroids or comet comae.  We also find a subtle narrow peak at 12.4--12.5~\micron{} in the comet Tempel~2 surface spectrum, and in the Tempel~1 coma spectrum.  Both of these features (at $\sim12.4$ and 15~\micron) remain unidentified.

Based on the comparisons between mid-infrared spectra of the comet surfaces and those of asteroids in the literature, the compositional similarity between comets and Jovian Trojans is strengthened.  Thus, these two classes of objects possibly have similar origins in the early Solar System.  However, comet comae show a wider diversity in the 10-\micron{} features than do our the set of comet and Trojan D-type surfaces presented here.  The anticipated \textit{James Webb Space Telescope} mission should be capable of addressing this latter point with spectra of more asteroids and bare comet nuclei.

\section*{Acknowledgments}

The authors thank the three referees for comments that improved this paper, J.~P.~Emery for sharing the Jovian Trojan spectra from \citet{emery06}, and C.~M.~Lisse for sharing the \spitzer{} spectra of comet Tempel~1 from \citet{lisse05-nucleus}.

M.S.P.K. and D.E.H. were supported for this work by the NASA (USA) Planetary Astronomy Program award NNX13AH67G.  C.E.W. was supported by Emerging Worlds Program award NNX16ADD33G.

This work is based in part on observations made with the \sst{}, which is operated by the Jet Propulsion Laboratory, California Institute of Technology under a contract with NASA (USA).  R.D.G., C.E.W., and M.S.P.K. were supported for these observations by NASA (USA) through Contract Nos.\ 1256406 and 1215746 issued by JPL/Caltech to the University of Minnesota.

This research made use of Astropy, a community-developed core Python package for Astronomy \citep{astropy13}.  This research also made use of Tiny Tim/Spitzer (STinyTim), developed by John Krist for the \spitzer{} Science Center.  The Center is managed by the California Institute of Technology under a contract with NASA (USA).

\bibliography{journals,references,extra}
\bibliographystyle{icarus}

\end{document}